\input psfig
\documentstyle[stacee_subfigure]{stacee_article}
\newcommand{\degs}{\mbox{\(^\circ \)}}
\begin {document}

\begin{center}
{\Large Prototype Test Results of the Solar Tower Atmospheric
Cherenkov Effect Experiment (STACEE)}
\bigskip

M.C. Chantell$^1$, D. Bhattacharya$^2$, 
C.E. Covault$^1$, M. Dragovan$^1$, R. Fernholz$^3$, D.T. Gregorich$^{4,5}$,  
D.S. Hanna$^3$, G.H. Marion$^1$,
R.A. Ong$^1$, S. Oser$^1$,
T.O. Tumer$^2$, D.A. Williams$^6$ 

\bigskip
$^1$ University of Chicago, Chicago IL 60637, USA

$^2$ University of California, Riverside, Riverside, CA 92521, USA

$^3$ McGill University, Montreal, Quebec H3A 2T8, Canada

$^4$ California State University, Los Angeles, Los Angeles, CA 90032, USA

$^5$ IPAC, California Institute of Technology, Pasadena, CA 91125, USA

$^6$ University of California, Santa Cruz, Santa Cruz, CA  95064, USA

\end{center}

\bigskip
\bigskip

\centerline{\em (Submitted to Nuclear Instruments 
and Methods in Physics Research A)}

\vspace{12mm}

\begin{abstract}
There are currently no experiments, either satellite or ground-based,
that are sensitive to astrophysical $\gamma$-rays in the energy range
between 20 and 250 GeV.  We are developing the Solar Tower Atmospheric 
Cherenkov Effect
Experiment (STACEE) to explore this energy range.
STACEE will use heliostat mirrors at a solar research facility to
collect Cherenkov light in extensive air showers produced by high
energy $\gamma$-rays.  Here we report on the results of on-site test work 
at the solar facility.
We demonstrate that the facility is suitable for use as an
astrophysical observatory, and using a full scale prototype of 
part of STACEE, we detect atmospheric Cherenkov radiation at energies
lower than any other experiment to date.  Based upon these results
we are confident 
that the eventual STACEE instrument will be capable of exploring the
$\gamma$-ray sky between 50 and 500 GeV with good sensitivity.
\end{abstract}

\bigskip


\newpage

\section{Introduction}

During the last few years the field of $\gamma$-ray astronomy has been
revolutionized by the discovery of over 100 point sources by the EGRET
satellite experiment with energies up to 20~GeV~\cite{egret}.  
At the same time, improvements in
ground-based telescopes using the
atmospheric Cherenkov technique have resulted in several
recent detections of point sources at energies above 250~GeV~\cite{review}.  
Still, most EGRET sources are not detected above 250~GeV.  For example,
of the large number of Active Galactic Nuclei (AGN) seen
by EGRET, only Markarian 421 has been detected by a ground-based 
instrument~\cite{punch}.  
This result implies a
spectral cutoff between 20 and 250~GeV, and may suggest that
high energy $\gamma$-rays are attenuated by photon-photon interactions
with the intergalactic infrared background~\cite{ir1,ir2}.  
A measurement of AGN
spectral cutoffs as a function of AGN redshift may be used to probe the 
infrared background, which is sensitive to details of galaxy formation
and dark matter composition~\cite{macminn}.  As another example, 
although it is generally
believed that supernova remnants (SNR) are important sites for cosmic ray
acceleration, to date no clear detection of $\gamma$-rays from SNR has been
made by ground-based instruments~\cite{buckley}.  
There is speculation that the $\gamma$-ray
spectra of SNR soften at energies above 20~GeV~\cite{DAV}, making it important to 
observe such objects at as low an energy as possible.  For these reasons,
the energy range between 20 and 250~GeV is expected to yield a wealth
of scientific discovery.  To date, however, it remains largely unexplored.
The existing satellite experiment (EGRET) has poor sensitivity above $\sim 10$ GeV
due to its limited collection area, while current ground-based experiments have energy 
thresholds above 250 GeV.

When high energy $\gamma$-rays or cosmic rays enter the
Earth's atmosphere, they interact and produce extensive air showers (EAS) of
highly relativistic charged particles.  These charged particles emit
Cherenkov radiation which forms a light pool $\sim100\,$m in radius at ground
level.  Atmospheric Cherenkov telescopes
operate by using large mirrors to collect this light and
focus it onto photomultiplier tube cameras.  The total amount
of Cherenkov light generated by an EAS is directly proportional
to the energy of the progenitor and thus, as one goes down in energy, the
density of Cherenkov photons on the ground decreases.  
The energy threshold of this type of instrument is
limited by the total mirror collection area. 
Larger mirror areas yield
lower energy thresholds.  Currently, the lowest energy
threshold obtained by an atmospheric Cherenkov telescope
is $\sim250$~GeV for the Whipple Observatory's
$10\,$m ($78.5\,$m$^{2}$ mirror area) telescope.

It has been recognized for some time that existing solar power plants represent
a potential resource for achieving lower energy thresholds due to their large
mirror areas~\cite{Danaher,tumer}.  Over the last few years,
we have been exploring the use of heliostat fields at such facilities,
with the goal of developing a new experiment called the Solar
Tower Atmospheric Cherenkov Effect Experiment (STACEE)~\cite{tumer2,astroparticle,
ong1}.  The goal of STACEE
is to explore the $\gamma$-ray sky in the energy range of $50$ GeV to $500$ GeV.
Our current plan is to make use of the National Solar Thermal Test Facility (NSTTF)
located in Albuquerque, NM.
STACEE will use 48 heliostats, spread out over a $150\,$m $\times \ 300\,$m 
area, to sample a large fraction of the Cherenkov light pool.
Each heliostat has a mirror area of $37.2\,$m$^{2}$, yielding a total
mirror collection area of $\sim1786\,$m$^{2}$.

In 1996, the STACEE collaboration designed and built a prototype
secondary telescope and camera for detecting atmospheric Cherenkov
showers using the heliostat field of the NSTTF
This prototype was installed
on the central receiver tower at the NSTTF for a series of tests in August
and October 1996.  The prototype included electronics and a data
readout system for eight photomultiplier tubes viewing eight
heliostats.

The purpose of these tests was to explore the feasibility of
establishing a $\gamma$-ray observatory at the Sandia site.
In particular, our goals for these
tests were to establish the following: 

\begin{itemize}
\item that
the environmental conditions at the Sandia site are
suitable for doing atmospheric Cherenkov astronomy,
\item that the mechanical and optical properties of
the heliostats are
of sufficient quality for astrophysical observations, and
\item that the STACEE concept is viable for a low threshold
ground-based $\gamma$-ray detector.
\end{itemize}

Our results on all three of these topics were excellent.  In the
remainder of this paper we elaborate our findings.

\section{Detector Concept}

The STACEE detector uses large steerable mirrors, called
heliostats, to collect Cherenkov light
from extensive air showers.  This light is reflected onto a secondary
mirror mounted on the central receiver tower.  The secondary mirror
images the light onto an array of photomultiplier tubes (PMTs) which
are mounted on a supporting structure.
Because the secondary mirror forms an image of the heliostat field in
its focal plane, each PMT may be positioned so that
it sees light from only one heliostat.
The PMTs together with the
supporting structure are referred to as the PMT camera. 
Figure~2.1 shows a schematic of the STACEE concept.
Complete details of the STACEE design concept can be
found in \cite{design}.

In our 1996 tests the signals from each PMT
were capacitively coupled, amplified, and delayed to
correct for the varying arrival time of the Cherenkov light at the
camera box.  PMT signals were combined to form a
fast trigger.  Analog to digital converters (ADCs), time to digital
converters (TDCs), and waveform digitizers recorded the pulse
amplitudes, times, and waveforms, respectively.  The data were read out
via CAMAC and GPIB by a 486-based PC.  
The prototype
tests used eight heliostats with eight PMTs.
Figure~2.2 shows a schematic of
the secondary telescope and camera and Figure~2.3 shows a schematic of
the electronics setup.  
Different sets of heliostats were used in the
August and October 1996 tests, as indicated in Figure~2.4.

\section{Results from Environmental Measurements}

The NSTTF is located on the grounds of Kirtland Air Force Base near
Albuquerque, New Mexico, at an altitude of $1700\,$m.  Environmental
conditions relevant to the STACEE project include:

\begin{itemize}
\item the local weather conditions, particularly the 
expected number of clear nights for astronomical observing,
\item the clarity of the atmosphere, particularly 
the impact of air pollution from nearby Albuquerque that might
attenuate Cherenkov light, and
\item the ambient light levels at night -- including the impact
of any extraneous light from Albuquerque.
\end{itemize}

\subsection{Local Weather Conditions}
Sandia National Laboratories are situated in the southwestern United
States in an area with a dry moderate climate.  
With the exception of
seasonal monsoons from July through August, this region receives little
precipitation.  Atmospheric levels of water vapor, which increases the
attenuation of light, are low, and the skies are usually very clear.
From meteorological records for Albuquerque (Figure 3.1) we estimate
that we will achieve an average of 4.1 hours of cloudless, moonless weather
per night during the Sept. through May observing season.
We estimate an annual duty cycle of
$\sim10\%$, which is typical for ground-based atmospheric Cherenkov
experiments at good observing sites.

\subsection{Atmospheric Clarity}

Atmospheric contaminants, such as aerosols and other pollutants, 
have the potential to attenuate the
Cherenkov light signal as it propagates through the atmosphere above
the Sandia site.  To determine the clarity of the local atmosphere at
Sandia we used stellar photometry to measure the optical transmission
of the atmosphere at the site as a function of atmospheric depth and wavelength.

The photometry data were collected over the course of a single night at
Sandia at a location approximately $20\,$ meters north of
the north edge of the heliostat field.  The photometer, as depicted
schematically in Figure 3.2, consists of a Hamamatsu photon-counting
photomultiplier tube (H3460-53) attached to the focus of a Celestron
8{\tt "} F/10 Schmidt Cassegrain Telescope (SCT).  
The PMT base housing contains amplifier and discriminator circuitry.
Output
pulses from the PMT are counted with a scaler which is read out
by a laptop PC.  The count rate from the PMT is directly proportional
to the photon flux falling on its photocathode.  The instrument
includes a set of standard Johnson UBV photometric filters that allows
measurements to be made in the Ultra-violet, Blue, and Visible
portions of the spectrum respectively.  Figure 3.3 shows the standard UBV
photometric response curves.  We note that the corrector
plate of the SCT is made of crown glass and hence does not transmit
below wavelengths of about $340\,$nm.  This cutoff affects measurements in
the U band, and is corrected for in the subsequent analysis of the
data.

We made repeated drift scans of four separate bright stars over a
five hour period.  Drift scans were done for each star using each
filter.  Each star was
scanned at five different zenith angles.  The scans included 
approximately 40 seconds of data before star transit in order
to measure the background light level.  Figure 3.4 shows the data from
a single scan.  The background and signal regions are clearly
separated, and the boundaries of the signal region are very sharp and
well defined.

The total atmospheric transmission factor for each of the UBV wavebands
is derived from the observations using standard photometric
techniques~\cite{photometry}. 
The observed flux of star light is found
from the data by calculating the average background rate from the data
before the the signal peak (see Figure 3.4) and subtracting that value
from the average count rate in the signal region of the drift scan,
taking care to avoid the regions where
the signal is turning on.  Error bars shown on subsequent
plots are the statistical errors associated with these averages.  

To find the unattenuated flux  of star light we use Bouguer's
law~\cite{photometry} which relates the apparent magnitude of a star 
($M_{o,\lambda}$) at the zenith as a function of wavelength to 
$(M_{z,\lambda})$ the magnitude as a function
of zenith angle $z$ and wavelength:

\begin{equation}
M_{z,\lambda} = A_{\lambda} \cdot \sec{z} + M_{o,\lambda}. 
\end{equation}

\noindent
Note that the logarithm of the photon flux is 
proportional to the stellar magnitude $M$.
To first order, $\sec{z}$ is proportional to the atmospheric depth
(sometimes called the airmass or overburden).  From a plot of 
$M_{z,\lambda}$ versus $\sec{z}$ the unattenuated stellar 
magnutide can be found by extrapolating to $\sec{z} = 0$ which
corresponds to an airmass of 0 (ie. no atmosphere). 
Figure 3.5 shows such a plot of the
data for the star Mu Andromeda with the V band filter.  

The atmospheric transmission is calculated as the ratio of the stellar
flux on the ground to the incident stellar flux at the top
of the atmosphere.  A weighted
average of results from all four observed stars yields the total
atmospheric attenuation for each wavelength band.

Figure 3.6 shows the atmospheric transmission derived from the
observations for the blue waveband as a function of zenith
angle.  The solid line shows the
fit of the data to the form:

\begin{equation}
T = a \exp(-b \sec(z)),
\end{equation}

\noindent
where $T$ is the atmospheric transmission factor, $a$ and 
$b$ are free parameters.
Table 1 gives the atmospheric
transmission at zenith for all three wavebands.

Figure 3.6 also shows the total transmission predicted from a model of
an ideal atmosphere.  This model assumes that the only light loss
mechanisms are Rayleigh and Mie scattering and absorption by oxygen
and naturally-occurring ozone~\cite{chantell}.   Instrumental effects
such as PMT quantum efficiency, wavelength dependent transmission of the
crown glass corrector plate, the reflectivity of the mirrors
in the telescope, and the appropriate transmission for the UBV filters have
all been folded in to the model.  
We can see
that the ideal atmosphere agrees with the observational data to within
3\% (see Table 1).  These measurements thereby demonstrate that the clarity of the
local atmosphere is not significantly affected by air pollution.
Table 1 also shows the predicted clarity for the Whipple Observatory for
comparison,
calculated assuming an ideal atmosphere; the Sandia and Whipple sites
have comparable atmospheric clarity.

\vspace{2mm}

\subsection{Ambient Background Light}

\indent
In addition to Cherenkov light from extensive air showers, the STACEE
detector is also sensitive to ambient light present at the site.  This
ambient light constitutes the background against which the Cherenkov
signal must be detected.  One source of ambient background is light
from the night sky viewed by the heliostats.  This includes
air glow, stars in the field-of-view of the
heliostats, and backscattered light from artificial sources on the
ground.  
Because the secondary mirror looks down into the heliostat field, a
second source of ambient light is light scattering off the ground
surrounding the heliostats.  This background can be minimized by
matching the aperture of each PMT to the size of the heliostat image
in the focal plane, so that the PMT sees only the surface of the
heliostat and not the ground around it.

Figure 3.7 illustrates the relative contributions of these backgrounds
when the heliostats are placed in different orientations.  The largest
background is seen when the heliostats are pointed to reflect light
arriving from zenith into the STACEE telescope.  Since we use Winston
cones to limit the field of view of each PMT to the area of a
single heliostat, we expect very little of the reflected light from
the ground to be picked up by the PMTs when the heliostats are pointed
towards the sky.  When the heliostats are turned so that they are
``edge on'' as seen by the STACEE telescope, the PMTs view only the
light reflecting off the ground around the heliostats, and the level
of the background drops by a factor of $\sim 1.5$.  Since the
background light level is reduced by taking the heliostats off the
night sky we are confident that the ambient light entering our PMTs 
is dominated by light from the night sky and not by light
produced by nearby Albuquerque.

We calculate the flux of background photons from the zenith by
measuring the currents in the PMTs.  The current from each PMT due to
background can be expressed as:

\begin{equation}
I = \Phi_{\mbox{\tiny bkg}} \ e \ G \ \Omega \ \epsilon \ A,
\end{equation}

\noindent
where $I$ is the PMT current in amperes, $\Phi_{\mbox{\tiny bkg}}$ is
the background photon flux, $e$ is the charge of the electron in Coulombs, $G$ is
the PMT gain, $\Omega$ is the solid angle in the sky viewed by each
heliostat, $\epsilon$ is the efficiency with which a photon incident
on the heliostat results in a photoelectron at the PMT, and $A$ is the
heliostat collection area.  We find that the flux
of photons is $4.3 \pm 0.9 \times 10^{12}$ ph/m$^2$/sec/sr.  For
comparison, the measured flux at a dark mountain site is $~2.0
\times 10^{12}$ ph/m$^2$/sec/sr~
\cite{lapalma}.

\section{Results from Heliostat Performance Tests}

\vspace{6mm}

The heliostat field at Sandia is designed to track the Sun and focus
its light onto the central receiver tower.  An individual heliostat
consists of 25 square mirror facets, each $1.22\,$m on a side, mounted on a
single altitude-azimuth mount. The surface of each facet is deformed
slightly using adjustable screws to obtain a parabolic reflective
surface.  The facets on each heliostat are aimed and focused to
produce the smallest possible image of the Sun on the receiver tower.

The elevation and azimuth positions of each heliostat are encoded with a
precision of 13 bits over 360\degs \  which corresponds to a precision
of 0.04\degs.  Pointing, tracking, and all other
aspects of heliostat operations are implemented by a central
controller.

To determine the suitability of the existing heliostats at the 
Sandia site for astronomical measurements, we measured  the following:

\begin{itemize}
\item pointing accuracy, 
\item tracking stability, 
\item focusing properties of the heliostats, and
\item reflectivity. 
\end{itemize}

Details for each of these
measurements are described below.

\subsection{Heliostat Pointing Accuracy}

Heliostat pointing accuracy is important to ensure that: 1) all
heliostats are viewing the same point in the sky, and 2) the collected
light is properly focused onto the secondary optic on the tower.

The pointing accuracy of the heliostats was determined by conducting
drift scans of several bright stars. During a single observation, all
heliostats were directed to observe a point seven minutes in right
ascension ahead of a bright star.  The heliostats were halted and the
star was allowed to drift through the field of view of the
heliostats while the PMT currents were recorded with a scanning ADC.
Figure 4.1 shows the light curves obtained for two heliostats during a
drift scan of the star Aldebaran.  The time of the peak current for
each PMT is found from the weighted average of the
background-subtracted light curve.  Table 2 lists the time, in seconds
from the start of the data run, at which the current reached a maximum
in each heliostat's light curve.

The data show that the selected heliostats were all aimed at a common
point to an accuracy better than $0.08\degs$
with a typical accuracy of $0.04\degs$ which is equal to the resolution
of the 13 bit heliostat position encoders.  Thus using star transit data
we can readily identify heliostats which are not properly
aligned and make corrections to fine-tune the heliostat pointing.

The light curves obtained from these drift scans can also be used to determine
the field of view for each heliostat.  
We define the field of view of a heliostat to be the full width half maximum
(FWHM) of its recorded light curve for a star transit.
From the known angular velocity of Aldebaran, we convert the
FWHM of each light curve into a field of view in degrees.
For the two light curves shown in Figure~4.1, we obtain an average heliostat
field of view of $0.7^\circ$.

\subsection{Tracking Stability}

The ability of a heliostat to maintain a celestial object centered in its 
field of view
as the object moves across the sky is referred to as the tracking
stability.  To examine the stability of the Sandia heliostats,
we tracked the bright star Betelgeuse for 18 minutes while recording
the PMT currents with scanning ADCs.  Figure 4.2 shows the light
curves for two of the heliostats.  The average current level
for each heliostat is stable to within a few percent over the
duration of the data run.  There is an obvious sawtooth modulation of
the currents with a period of $\sim 30$ sec. The sawtooth pattern
seen in the lower plot of Figure 4.2 results from the heliostat 
repeatedly being
brought on target (the point of maximum current in the PMT) and then
slowly drifting off target.  The distance through which the target
drifts before the heliostat is commanded to update its position is
equal to the single bit resolution of the heliostat position encoders.  
Knowing the apparent
motion of Betelgeuse across the sky, we find
that the periodicity of the observed sawtooth pattern corresponds
precisely to the period expected from the encoder resoultion.

Examination of data from another heliostat shown in the upper plot
of Figure 4.2
reveals the same type of sawtooth pattern but with the orientation of
the ``teeth'' in the opposite sense.  In this case the heliostat was
tracking a position slightly ahead of the
target star. Thus the current dropped suddenly when the heliostat
moved to reacquire the target and then slowly rose as the star drifted
back into the field of view of the heliostat.  These results demonstrate how star
tracking can be used as a powerful diagnostic tool for evaluating the pointing
bias of individual heliostats.  We plan to conduct regular
star tracking runs as a means of monitoring the heliostat tracking
stability.  Through the use of these runs and stellar drift scans,
we expect to be able to
correct the pointing biases of all 48 heliostats to within $~0.04\degs$.  
This accuracy is more than adequate, given STACEE's expected angular resolution of
$\sim0.2\degs$.

\subsection{Optical Properties of the Heliostats}

The critical optical property of the heliostats is their focusing.  In
order to maximize light collection it is important that the
reflected light from a heliostat be focused onto as small an area on the central
tower as possible.  The minimum acceptable size for the secondary
mirrors is determined by the projected heliostat spot size.

The heliostat optics were evaluated by imaging the Sun onto the front
face of the tower and recording the images with a CCD camera.  Fifteen
of the sixteen heliostats used in the two prototype tests were tested
in this manner; we were unable to obtain data for one heliostat
due to cloudy weather.  Figure 4.3 shows a typical
heliostat image of the Sun.  From these images we have determined the
average FWHM heliostat spot size to be less than $2.0\,$m
diameter.  Since the Sun has an angular extent of {0.5\degs}, the size
of its image closely matches the expected size of a 50~GeV $\gamma$-ray
air shower.  Table 3 lists the Sun spot sizes for all fifteen
heliostats tested.  Figure~4.4 shows the FWHM of the
spots as a function of the 
distance between each heliostat and the target. 
These data indicate a regular trend
towards a tighter concentration of light for heliostats closer to the
central tower.  We can use this trend to 
predict the light profile expected at the target from
any heliostat in the field.

\subsection{Heliostat Reflectivity}

To estimate the total light collection efficiency of the STACEE
optical system, it is necessary to know the reflectivity of the
heliostats as a function of wavelength.  The heliostats at Sandia are
back-surface silvered glass approximately 3~mm thick.  Because the
glass is opaque to UV light, a significant amount of the Cherenkov
flux from air showers is unavoidably lost at the heliostat.  Measurements
performed by Sandia personnel place the average heliostat reflectivity
at $80\%$ for visible light. To measure the reflectivity as a function
of wavelength, we designed and constructed a custom 
reflectometer which was used in the
field. The reflectometer uses a collimated light source and a high
quality PMT to measure light reflected from the surface of a mirror.
The instrument is calibrated 
using a standard mirror of known reflectivity.
Narrow bandpass filters allow the measurement of reflectivity as a
function of wavelength. The results are shown in Figure 4.5.  These
measurements demonstrate that the overall reflectivity is $\sim85\%$, and
that the UV cutoff occurs near $330\,$nm.  Since the heliostats are already
over 20 years old, and have maintained a high degree of reflectivity,
we do not expect their optical performance to significantly degrade
over time scales relevant to the STACEE project.

\section{Prototype Performance Results}

In addition to establishing the suitability of the Sandia site for
astronomical observations, we also built and tested a fully functional
STACEE prototype (See Figure~2.2).  The prototype included a secondary
telescope, PMT camera, and data acquisition electronics for 8 channels.  In
both the August and October tests, we ran the prototype STACEE
experiment and investigated the following: 

\begin{itemize}
\item optical characteristics of the secondary mirrors, 
\item performance of the analog trigger, 
\item performance of the digital trigger and coincident trigger rates,
\item performance of waveform digitizers.
\item energy threshold obtainable, and 
\end{itemize}

\noindent
The secondary telescope and camera were installed on one of the test
bays of the central receiver tower.  This bay is 10$\, '$ deep,
35$\, '$ wide, and 160$\, '$ above the heliostat field, and looks out on
the field to the north.	

\subsection{Optical performance of the secondary mirrors}

A critical component to the STACEE telescope is the secondary mirror
that collects the Cherenkov light from the heliostats and reflects
it on to the PMTs.  We have explored 
several design approaches for developing large, highly reflective
mirrors at low cost. Two distinct mirror technologies have been tested
in the field. During the August test we used a $3\,$m diameter
multifaceted secondary made with a stretched aluminum membrane mirror
technology. During the October test, we used a $1.8\,$m 
diameter back-silvered slumped glass mirror.

\subsubsection{Stretched membrane mirrors}

For the August test we used a secondary mirror with seven facets
made of electro-polished  stretched aluminum sheet.  
The individual facets had spherical
curvature and were aligned at
night by directing a high power searchlight beam onto a heliostat
which reflected the light onto the secondary mirror.  
Each facet produced an optical spot of approximately $4\,$cm diameter on
a target at the focal plane.  To co-align the facets, six facets were
covered with black cloth while the seventh facet was adjusted using a
three point turnbuckle arrangement.  This procedure was repeated until
the optical spots from all seven facets were aligned.
The aggregate alignment of the secondary system was cross-checked by
tracking bright celestial objects (such as the planet Jupiter) and
projecting the collected light pattern onto a white lucite board
mounted at the focal plane.

\subsubsection{Slumped glass secondary mirror}

For the October test we used a slumped glass parabolic mirror
($1.8\,$m diameter) as the secondary optic.  Using a single mirror
eliminates the need to align multiple facets. 
The slumped mirror also provided superior optical surface quality,
enough so that the images of the heliostats were clearly visible at the
focal plane during daylight.  Figure 5.1 shows the images of eight
selected heliostats projected onto a white placard at the focal plane.
Each of the eight heliostats is cleanly imaged by the secondary, with
virtually no optical overlap. 
The optical spot size for this mirror is less than 1 cm.

\subsubsection{Optical crosstalk}

For an ideal secondary mirror system, all of the reflected light from
each heliostat in the field will be focused onto the collecting
aperture of a single PMT.  However, optical aberrations, imperfections
in mirror surface quality, and facet misalignment may combine to
defocus the image of the heliostats at the focal plane, causing a 
small amount of light from one heliostat to
enter another heliostat's PMT.
We refer to this as optical crosstalk.  We measure the amount of
crosstalk using Cherenkov light signals from cosmic ray air shower data.  
For each event,
time-to-digital converters (TDCs) were used to record the relative
arrival times of the Cherenkov light pulses at each PMT. Since the
time-of-flight depends upon the unique distance between each heliostat
and the secondary, photons that were reflected into the wrong PMT will
be out of time with other photons collected from the same Cherenkov
event.  Analysis of the TDC differences between all combinations of
adjacent channels show that the maximum level of crosstalk is
approximately $1\%$.  There is no evidence for crosstalk between
non-adjacent heliostats.  The largest amount of optical crosstalk
occurs for physically adjacent channels in the same heliostat row.
(Note that with the application of cuts on TDC times and a strategic
selection of heliostat locations, optical crosstalk is expected to be
negligible for the STACEE experiment). 

\subsubsection{Summary of Mirror Tests}

Measurements made with two different mirror technologies show both are
suitable for use as the secondary optic.  The image sizes
are small enough to be contained by 5{\tt "} diameter Winston cones.
Optical crosstalk is small and easily rejected off-line by exploiting
the time-of-flight differences between different heliostats. 

The single-piece slumped glass secondary gives an optically superior
performance over the multi-faceted aluminum secondary, and is also
easier to mount and align.  Therefore this technology is currently
considered the most promising for STACEE.
The STACEE group continues to develop and improve mirror
performance using both technologies.

\subsection{Analog vs. Digital Trigger}

Atmospheric Cherenkov telescopes are typically triggered by
requiring a minimum number of PMTs to exceed a preset threshold within
a small ($\sim 10$ ns) time interval.  Two different methods exist for
implementing such a trigger.  An {\em analog}\/ trigger takes the analog
signals from each PMT and sums them.  This summed signal is sent
to a discriminator which produces a trigger if the summed
signal exceeds a preset threshold.  In a {\em digital}\/ trigger each
channel is individually discriminated and a trigger is generated when
some minimum number of discriminated channels fire within a specified
time interval.  In deciding on a suitable trigger scheme the following
characteristics must be considered:

\begin{itemize}
\item sensitivity to accidental 
triggers resulting from fluctuations in the NSB and afterpulsing in
the PMTs,
\item sensitivity to local phenomena such as single energetic muons 
passing close to an individual heliostat, and
\item achievable energy threshold.
\end{itemize}

\subsubsection{Analog Trigger}
The primary advantage of an analog trigger is that it makes full use
of the signal-to-noise improvement that comes from increasing mirror
collection area.  It has been shown \cite{weekes} that
the achievable energy threshold of an atmospheric Cherenkov telescope goes as:
\begin{equation}
E_{th} \propto \sqrt{{\Phi_{bkg} \Omega \tau}\over{\epsilon A}}\ \ .
\end{equation}
\noindent
Here, $\Phi_{bkg}$ is the flux of night sky photons, $\Omega$ is the field 
of view,
$\tau$ is the trigger gate width, $A$ is the mirror collection area,
and $\epsilon$ is the efficiency for converting a photon striking a
heliostat into a photoelectron in its PMT.  Of these quantities,
$\Omega$ and $\tau$ are constrained by the physics of air showers and
$\epsilon$ is constrained by technology.  The most direct way to lower
the energy threshold of an atmospheric Cherenkov telescope
is to increase the effective mirror
area.  Since an analog trigger takes full advantage of information
contained in the time structure of the PMT pulses, in principle it
provides the lowest possible energy threshold.

We tested this form of trigger by pointing all eight heliostats to
different parts of the sky, near the zenith, so that their 
fields of view did
not overlap.  In this configuration, triggers result only from unwanted
sources of background, and not from Cherenkov radiation seen in common
by a number of heliostats.  We measured the trigger rate as
a function of discriminator threshold.  The results are shown in
Figure 5.2.  The data show a clear spectral break near a discriminator
threshold of $\sim 200\,$mV.  If the camera were triggering only on
fluctuations of the background photon flux we would expect to see a
single, steeply falling spectrum.  To understand the presence of the
second spectral component at high thresholds, we use Monte Carlo
techniques, together with a model for the optical throughput of the
STACEE prototype, to estimate the expected trigger rates resulting from
single energetic muons passing near individual heliostats and for
single heliostats triggering on small cosmic ray air showers.

To estimate the expected rates, we determine the number of
photoelectrons necessary to cause a trigger as a function of discriminator
threshold from the known gain of the PMTs ($\sim 0.6 \times 10^6$), 
the average PMT pulse
width ($\sim 10\,$ns FWHM), 
and the measured losses due to
cable attenuation ($\sim 0.3$).
The conversion between photoelectron signal and
discriminator threshold is estimated to be 0.28~photoelectrons/mV.  
Using the average photon density on the ground, as determined by
Monte Carlo, we estimate the energy 
threshold for protons and single muons as a function of discriminator
threshold.

The trigger rate for a single heliostat can then be found from:
 
\begin{equation}
R = \Phi \ A_{\mbox{\footnotesize eff}} \  \Omega \  \epsilon,
\end{equation}
 
\noindent
where $R$ is the trigger rate, $\Phi$ is the flux of Cherenkov-generating background
particles (cosmic ray primaries or muons),
$A_{\mbox{\footnotesize eff}}$ is the effective collection area for a single
heliostat, $\Omega$ is the solid angle viewed by a heliostat, and
$\epsilon$ is the efficiency for converting photons into
photoelectrons.

For cosmic ray air showers we use an effective all-particle cosmic ray
flux which has been corrected for the varying Cherenkov yield as a
function of composition \cite{astroparticle}:

\begin{equation}
\Phi_{cr} = 9.1 \times 10^{-6} (\frac{E}{1000~GeV})^{-1.67}\mbox{ showers/cm$^{2}$/s/sr},
\end{equation}

\noindent
where $E$ is the effective energy threshold in GeV.  For local muons we use the 
flux of muons above an energy of 
7~ GeV \cite{alkofer}: 

\begin{equation}
\Phi_{\mu} = 1.52 \times 10^{-3} E^{-1.24} \mbox{ $\mu$/cm$^{2}$/s/sr} ,
\end{equation}

\noindent where $E$ is taken to be 7 GeV which is approximately the energy below which 
Cherenkov production from muons ceases.  From drift scan data (see Section 4.1)
we find that $\Omega$ 
for a single heliostat
is $\sim 1.2\times 10^{-4}\,$sr.  From Monte Carlo simulations of proton showers
and individual muons we determine 
$A_{\mbox{\footnotesize eff}}$ to be $15400\,$m$^2$ for protons and
$154\,$m$^2$ for muons.
Since all eight heliostats are triggering independently on different
parts of the sky the total trigger rate for all eight heliostats
combined is eight times the rate for a single heliostat (cosmic ray triggers
+ $\mu$ triggers).  The predicted rates are in good agreement with the
measured rates for pulses greater than 200 mV, as shown in Figure~5.2.
This analysis demonstrates that the simple analog trigger is
sensitive to single heliostat triggers resulting from single muons and
small cosmic ray showers.

\subsubsection{Digital Trigger}

To test the rate of background triggers for a digital trigger we
measured the coincidence rates for different combinations of PMT
threshold and multiplicity.  As for the analog trigger we performed these
measurements with the heliostats pointed at different parts of the sky
to determine the rate of unwanted background triggers.
Figure 5.3 shows the rate versus threshold curves for three different
multiplicity requirements.
Note that in contrast to the analog trigger, here we do not
see a break in the spectrum, which indicates
that the observed rates are due only to
fluctuations in the NSB and not a secondary source of unwanted triggers.

\subsubsection{Comparison of Analog and Digital Triggers}

From the background rate versus threshold curves for the two trigger
types we see that the analog trigger is sensitive not only to
fluctuations of the background photon flux, but also to single
heliostat triggers from cosmic ray air showers and local muons.
The digital trigger appears to be sensitive only to
the background photon flux.  
The analog trigger is
sensitive to large amplitude signals in any individual channel whereas
the digital trigger demands that the light from an event be spread out
over several heliostats.  Since we wish to trigger selectively on
Cherenkov light from $\gamma$-ray initiated air showers, which will
uniformly illuminate heliostats over a $100\,$m radius, a digital
trigger or a modified analog trigger that incorporates some
multiplicity requirement is best suited to our needs.  

\subsection{Coincidence Event Studies}

To fully understand the performance of the digital trigger, we studied
the rate of coincident events when the heliostats
were aimed to a common point in the sky.
Using these measurements, we:

\begin{itemize}
\item established that the system was triggering on cosmic ray showers, 
\item investigated the zenith angle dependence,
\item investigated the rates as a function of heliostat canting angle, and
\item determined the energy threshold for the prototype telescope.
\end{itemize}

\subsubsection{Cosmic Ray Triggers}

To establish that the STACEE prototype is triggering on genuine air
shower events we took coincident data with the heliostats
in two separate viewing configurations: 
all heliostats viewing separate areas of the sky 
near the zenith (random mode), and all heliostats simultaneously viewing the 
zenith (coincident mode).  A
digital trigger condition of any four out of eight (4/8)
PMTs at a PMT threshold of $31\,$mV was
used, and PMT pulse height and timing information were recorded.
When the heliostats were in random mode
we expected no coincidences due to cosmic-ray air showers.
In this configuration we observed an event rate of 0.2 Hz.  When the
heliostats were placed in coincident mode the event rate increased to
5.1 Hz.  This indicates that our rate of genuine coincident events due
to cosmic ray air showers was 4.9 Hz.

To further establish that the STACEE prototype was triggering on cosmic
ray air showers we examine the PMT pulse height distribution from the
data taken with the heliostats viewing the zenith in coincident mode.
Since
the field of view of each heliostat is well matched to the angular extent
of an air shower, the total light collected by a channel 
is directly related
to the energy of the air shower progenitor.  
We expect that the spectrum of observed
pulse heights in the individual channels should match the differential
cosmic ray energy spectrum at 100~GeV.  In this energy range, the cosmic ray
spectrum has a power-law dependence on energy, $dN/dE \propto
E^{\alpha}$, where $\alpha$ is $-2.65$.  We analyzed 35 minutes of data
taken with the heliostats observing the zenith in coincidence mode. 
The digitized pulse
heights of all channels exceeding threshold were combined to build a
pulse height spectrum (Figure 5.4).  The rising part of the observed
spectrum is due to a convolution of trigger inefficiencies for small
pulses and fluctuations in shower brightness at energies near
threshold.  The trigger becomes fully efficient for events with pulse
heights above 60 digital counts. A fit to the falling edge of the
spectrum
yields a power law index $\alpha$ of $-2.7\pm0.1$ (statistical error
only), which agrees well
with the known differential index of the cosmic ray
spectrum in this energy range. Thus, we are
confident that we were in fact triggering on Cherenkov light from
cosmic ray air showers.

\subsubsection{Zenith angle effects on system trigger}

Because of the celestial motion of sources across the sky, $\gamma$-ray
observations must be conducted over a range of zenith angles.  The
effective energy threshold of an atmospheric Cherenkov telescope
varies with zenith angle due to
the increased atmospheric overburden.  As the overburden increases air
showers develop further away from the ground, resulting in: (1) an
increase in the attenuation of the Cherenkov light, and (2) a
spread in the Cherenkov light pool over a larger area, reducing
the photon density on the ground.  Both of these effects are
proportional to the increase in atmospheric depth with zenith angle and
therefore,
we expect the event rate to decrease with zenith angle.
Figure 5.5 shows
the observed event rates as a function of zenith angle, which are well
fit by a cos$^{2}(\theta)$ function.  Note that the event rates fall
off slowly out to a zenith angle of $30\degs$, after which they
fall more steeply.  
In order to maximize sensitivity,
STACEE will concentrate on observing sources within $30\degs$ of the zenith.

\subsubsection{Effect of heliostat canting on event rates}

Since the heliostats have small fields of view, 
maximum trigger sensitivity is
obtained when the heliostats are pointed to the shower interaction region 
(approximately $10\,$km above sea level).
The interaction region is defined as the
location where most of the Cherenkov light is generated for events
that land on the geometric center of the heliostat field.  Tracking
the heliostats to observe the interaction region, rather than towards
the source at infinity, requires a small adjustment in the pointing
angle of each individual heliostat.  This adjustment is called
``canting'' because each heliostat is
canted slightly inwards so as to
observe the interaction region.

Figure 5.6 illustrates the application of correct
heliostat canting.  Figure 5.6a shows the fields of view of the heliostats in
the case where all heliostats are tracking parallel in the direction
of the source at infinity.  In this case, coverage of the event track
is incomplete.  Figure 5.6b shows the heliostats correctly canted so
that each heliostat will collect light from the full longitudinal
development of the air shower.  The interaction region corresponds to
the point of shower maximum in the development of an air shower and occurs
at an atmospheric depth of ~270 g/cm$^2$ ($\sim10$\,km altitude) for showers initiated by
50~GeV $\gamma$-rays.  Figure 5.7 shows the effect of changing the
heliostat canting angle on the trigger rate.
The distribution is well fit by a Gaussian with a
mean of $0.15\degs$, corresponding to a depth of shower maximum
consistent with the expected value for cosmic rays at energies of a few
hundred GeV which compose the bulk of the event triggers (see section 5.3.5).  

\subsubsection {Performance of Waveform Digitizers}

A key goal of the on-site tests was to investigate the suitability of
waveform digitizers for pulse timing and amplitude measurement.  As
part of the electronics setup, we used Tektronix 644A and Tektronix
740A digital oscilloscopes to digitize the PMT waveforms for all eight
phototubes.  The Tektronix 644A scope had a sampling rate of 2 GSample/sec,
while the 740A had a sampling rate of 500 MSample/sec.  Sampling at 1 GSample/s
was simulated with the 644A by averaging consecutive data points. 
Each scope had a
dynamic range of 8 bits.  The digitized waveforms were read out over a
GPIB interface and saved to disk.  The GPIB interface readout
introduced a significant deadtime (nearly 0.4 sec per event), 
and thus the waveform
digitizers were only read out on selected runs.  

For vertical showers, the maximum difference in shower arrival times
at two adjacent heliostats spaced $10\,$m
apart is about 0.6 ns.  (Note that here we refer only to shower arrival times
at the heliostats in the field and not the difference in shower arrival
times at the secondary telescope on the tower).
Thus, the spread in photon arrival times is
smaller than the expected measurement precision of $\sim 1$ ns.
Hence, measurements of Cherenkov showers taken at zenith serve as a
calibration beam for timing measurements, and the spread in the difference of
arrival times between two adjacent heliostats is dominated by the
timing resolution of the electronics.

The expected location of the Cherenkov pulse in each digitized
waveform can be determined from the known detector geometries and
cable delays.  Using data within a 15 ns window centered around the
expected pulse location, the arrival time of the Cherenkov pulse was
defined to be the centroid of all data points exceeding a threshold of
30 mV.  We find that this technique yields better fitted times with a smaller
spread than other methods, including a Gaussian fit to the pulse shape.

To determine the timing resolution, we calculated the quantity:
\begin{equation}
\Delta t = t_{2} - \frac{t_{1} + t_{3}}{2},
\end{equation}
where, $t_{1}$, $t_{2}$, and $t_{3}$ are the measured pulse arrival
times at three consecutive heliostats in the same row. 

We calculate the RMS spread in $\Delta t$, $\sigma(\Delta t)$,
for timing measurements made with the
waveform digitizers and for measurements made using conventional
TDCs on the same data.  Assuming independent measurements, the timing
resolution, $\sigma_{t}$, is equal to:

\begin{equation}
\sigma_{t} = \frac{\sigma(\Delta t)}{\sqrt{\frac{3}{2}}}.
\end{equation}

We find that the waveform digitizers give a
narrower spread in $\Delta t$ than the TDCs for all pulseheights (See
Table 4).  In addition, the digitizers allow us to fit pulse arrival
times for many pulses which were too small to trigger the TDC.  Thus,
using the waveform digitizers, we are able to reconstruct arrival
times for events which cannot be reconstructed by TDCs alone.

We also studied the applicability of waveform digitizers for pulse
charge measurement.  By summing the digitized voltages recorded within
an integration region around the pulse arrival time, we determined the
total charge in the pulse. We find that the waveform digitizers give
charge measurements which agree with those found from conventional
ADCs.

\subsubsection{Energy Threshold}

We estimate the effective energy threshold 
for cosmic rays and $\gamma$-rays
of the eight heliostat configuration used during the October test.
We calculate the energy threshold for cosmic rays from the measured
event rate while observing the zenith.  The observed rate is related to
the energy threshold by:

\begin{equation}
R = \Phi_{cr} \ A_{\mbox{\footnotesize eff}} \ \Omega,
\end{equation}

\noindent 
where $R$ is the observed rate, $\Phi_{cr}$ is the flux of cosmic ray showers, 
$A_{\mbox{\footnotesize eff}}$ is the effective collection area, and
$\Omega$ is the solid angle acceptance of the instrument.  For the flux
of cosmic rays we use the equivalent proton flux given by Equation 6.

The maximum observed trigger rate was 5.1~Hz, using a digital trigger
with a multiplicity of 4/8 PMTs and a tube threshold of 31~mV.  The
accidental trigger rate in this configuration was 0.2~Hz, and was
subtracted from the total rate to yield 4.9~Hz.

The effective collection area and solid angle acceptance
for proton-induced air showers are calculated from Monte Carlo simulations.
We simulate proton air showers using 
MOCCA \cite{hillas}, a well-established air shower Monte Carlo.  The simulated
showers are then processed through a detailed ray tracing simulation
of the Sandia heliostat field.
This simulation takes into account the measured focusing properties of the 
heliostats and secondary mirror, reflectivity losses, and ambient
background light levels.  Proton
showers were simulated at energies from 50~GeV to 2~TeV 
and weighted according to
the cosmic ray spectrum.  The simulated showers were then dropped 
randomly onto the
heliostat field over a 100~m radius area around the heliostats, 
and the incident angle of the showers were randomly varied 
within $1.0\degs$ of zenith.  The effective
collection area calculated
this way is:

\begin{equation}
A_{\mbox{\footnotesize eff}}=1.9 \pm 0.7 \times 10^{8} \textrm{cm}^{2},
\end{equation}

\noindent
while the solid angle acceptance is:

\begin{equation}
\Omega=3.7 \pm 0.2 \times 10^{-4} \textrm{sr.}
\end{equation}

\noindent
Note that the solid angle acceptance of the eight heliostats in coincidence
is larger than the individual heliostat acceptances.  This is due to the fact
that the system is triggering on air showers which have an angular extent of
$~0.5\degs$.

Applying Equation 8 together with Equation 6 yields a cosmic ray energy
threshold of:

\begin{equation}
E_{\mbox{\footnotesize cr}} = 295^{+96}_{-50} \ \textrm{GeV.}  
\end{equation}

An alternative way of estimating the effective energy threshold
is to compare the measured photoelectron
yields to the expected photon densities on the ground as a function of
primary energy.  We use the MOCCA simulation to determine the average
photon density on the ground within 100~m of the shower core, as a
function of energy (see Figure~5.8).

Figure 5.4 shows the pulse height spectrum built up from all PMT pulses 
which exceeded trigger threshold for observations made at zenith. 
We define the trigger threshold 
pulse height as the pulse height above which the integrated area of the
power law fit equals the area under the data curve (ie. the trigger rate
predicted by the fit equals the observed rate).  This corresponds
to the minimum pulse height needed to trigger a PMT.
The trigger pulse
height is found to be: 
\begin{equation}
P_{\mbox{\footnotesize trig}} = 33^{+8}_{-6} \ \textrm{digital counts,}
\end{equation}
where the uncertainty is due to the
uncertainty in the spectral index of the fit.  Knowing 
the rate at which charge is converted to digital counts
in the ADC 
($Q = 0.25\,$pC/digital count), the PMT gain ($G_{\mbox{\footnotesize pmt}} = 
6.3 \pm 1.3 \times 10^5$) and the gain of
the pre-amp ($G_{\mbox{\footnotesize amp}} = \times 10$) we can relate
the equivalent number of photoelectrons ($N_{pe}$)
at trigger threshold to the trigger threshold pulse height by:

\begin{equation}
P_{\mbox{\footnotesize trig}} = \frac{N_{pe} \ G_{\mbox{\footnotesize pmt}}
\ G_{\mbox{\footnotesize amp}} \ e}{Q},
\end{equation}

\noindent
where $e$ is the charge of an electron.  From this we find:

\begin{equation}
N_{pe} = 8.3^{+2.0}_{-1.5} \ \textrm{photoelectrons.}
\end{equation}

To convert from photoelectrons at the photocathode to photons striking the heliostat 
($N_{\mbox{\footnotesize phot}}$) we divide by
the average PMT 
quantum efficiency (0.21, a convolution of the Cherenkov spectrum on
the ground and the wavelength-dependent quantum efficiency of the PMT), 
and by the collection efficiencies due to
heliostat reflectivity (0.8), secondary mirror reflectivity (0.85), 
heliostat focusing efficiency on the secondary (0.8), 
secondary focusing efficiency on the PMT
winston cone (0.85), and the throughput of the Winston cone (0.9).  This
gives the number of photons incident on a heliostat: 

\begin{equation}
N_{\mbox{\footnotesize phot}} = 95^{+27}_{-23} \ \textrm{photons,}
\end{equation}

\noindent
where a 15\% systematic error has been included to account for uncertainties in
the efficencies and reflectivities.

Dividing by the
projected area of a heliostat (29~m$^2$) gives the photon density ($\rho$)
on the ground necessary to trigger a heliostat channel:

\begin{equation}
\rho = 3.3^{+0.9}_{-0.8} \ \textrm{photons/m$^2$}.
\end{equation}

\noindent
From Figure 5.8 we can use $\rho$ to estimate the lowest energy $\gamma$-ray
that will produce a high enough photon density to trigger the experiment.
We find the energy threshold, for $\gamma$-rays from a source at the zenith, 
to be:

\begin{equation}
E_{\mbox{\footnotesize $\gamma$}} = 74^{+17}_{-14} \ \textrm{GeV}.
\end{equation}

\noindent
For cosmic rays, the same photon density corresponds to an energy
threshold of:
\begin{equation}
E_{\mbox{\footnotesize cr}} = 385^{+55}_{-60} \ \textrm{GeV.}  
\end{equation}
This energy threshold agrees within error with that obtained from the
rate calculation.

\section{Summary}

We are developing a novel atmospheric Cherenkov experiment called
STACEE that will use solar heliostat mirrors at the National Solar 
Thermal Test Facility (NSTTF).
Two on-site tests have been conducted.  These tests have demonstrated
that the heliostat
field at the NSTTF is suitable for use as the primary optical
component of an astrophysical $\gamma$-ray 
detector.  Weather and atmospheric conditions at the site compare favorably
to conditions at existing observatories.  Measurements of the
ambient background light levels show that the close proximity of the
site to Albuquerque will not adversely effect the performance of STACEE.
The mechanical and optical performance of the heliostat field is found
to be excellent, in most cases greatly exceeding our requirements.

The NSTTF is a scientific research facility, and so possesses an excellent
support infrastructure.  This infrastructure includes heavy lift capability,
a high bay, a complete machine shop, and other facilities necessary for 
developing and running an experiment such as STACEE.  We feel that the 
NSTTF site is well suited in all respects for astrophysical
observations.

The STACEE prototype's secondary telescope, camera, and associated electronics
also performed to expectations.  The secondary optics produce well-focused
heliostat images that are completely contained by the PMT Winston cones.
Measurements of cosmic ray air showers show that the system is stable and
responds as expected.  The measurements indicate that our prototype
detector had a $\gamma$-ray energy threshold below 100 GeV, and that the STACEE
concept can successfully obtain lower energy thresholds than existing 
ground-based techniques.  

\begin{center}
\textbf{Acknowledgments}
\end{center}

We are grateful to the Physics Division of Los Alamos National Laboratory
for loans of electronics equipment.
We thank the staff at the NSTTF and in particular, 
acknowledge the contributions made by 
J.M. Chavez, R.M. Edgar, C.M. Ghanbari, D. Johnson, J.J. Kelton,
L. Killian, and R. Tucker.
We also acknowledge the assistance of E. Pod,
the engineering staff of the Yerkes Observatory, 
the personnel of the McGill Physics Department shop.
We wish to thank M. Cresti for the use of a 1.8m mirror,
and J. Carlstrom of the University of Chicago for the loan of a digital
oscilloscope.
This work was supported by the National Science Foundation,
the Institute of Particle Physics of Canada, the Natural  
Sciences and Engineering Research Council, and the
California Space Institute.
TOT wishes to acknowledge the
support of the University of California, Riverside, Vice Chancellor of
Research and College of Natural and Agricultural Sciences.
CEC wishes to acknowledge support from the Louis Block Fund
of the University of Chicago. 
RAO wishes to acknowledge the support of the Grainger Foundation
and the Physical Sciences Division and the Enrico Fermi Institute
of the University of Chicago.

\newpage

\begin{table}[h]
\begin{center}
\begin{tabular}{|c|c|c|c|}
\hline Waveband & U & B & V \\\hline
Measured (Sandia) & 0.548 & 0.707 & 0.813 \\\hline
Model Atmosphere (alt.=1700m) & 0.562 & 0.726 & 0.841 \\\hline
Predicted Whipple (alt.=2300m) & 0.590 & 0.760 & 0.861\\\hline
\end{tabular}
\caption{Atmospheric transmission factors for star light from the
zenith.  Note that $1700\,$m corresponds to the altitude of
Sandia.}
\end{center}
\end{table}

\newpage

\begin{table}[h]
\begin{center}
\begin{tabular}{|c|c|c|} \hline
Heliostat ID & Transit Time (s) & $\Delta\degs$ \\\hline
8E3 & 433 & 0.05 \\
8E4 & 433 & 0.05 \\
8E5 & 402 & -0.08 \\
8E6 & 418 & -0.01 \\
10E4 & 437 & 0.07 \\
10E5 & 421 & 0.00 \\
10E6 & 412 & -0.04 \\
10E7 & 413 & -0.03 \\\hline
\end{tabular}
\caption{Star transit times for each heliostat for the drift scan of Aldebaran.
The angular separation ($\Delta\degs$) of each heliostat, relative to the
average of the eight transit times, is shown.}
\end{center}
\end{table}

\newpage

\begin{table}[h]
\begin{center}
\begin{tabular}{|c|c|c|ccc|c|} \hline
 &  Distance &      & \multicolumn{3}{c|}{Radius (m)} & Luminance within\\
Helio &    to target & FWHM & \multicolumn{3}{c|}{to contain luminance} &
2.0 m diameter \\
ID & (m) & (m) & 50\% & 80\% & 90\% & (percent) \\ \hline
8E3 &  103.9 & 0.93 & 0.40 & 0.63 & 0.82 & 94 \\
8E4 &  106.9 & 0.94 & 0.41 & 0.64 & 0.83 & 94\\
8E5 &  110.4 & 1.27 & 0.61 & 0.97 & 1.25 & 82\\
8E6 &  114.6 & 0.99 & 0.44 & 0.68 & 0.85 & 94\\
10E4 & 131.4 & 1.04 & 0.45 & 0.70 & 0.87 & 94\\
10E5 & 134.3 & 1.10 & 0.51 & 0.80 & 0.99 & 90\\
10E6 & 137.8 & 1.08 & 0.50 & 0.79 & 0.99 & 90\\
10E7 & 141.9 & 1.14 & 0.53 & 0.84 & 1.05 & 88\\
12W1 & 158.8 & 1.21 & 0.57 & 0.88 & 1.10 & 86\\
12E1 & 158.8 & 1.29 & 0.63 & 1.01 & 1.33 & 80\\
12E2 & 159.4 & 1.24 & 0.61 & 1.04 & 1.50 & 78\\
14E1 & 198.2 & 1.67 & 0.81 & 1.25 & 1.53 & 65\\
14E2 & 198.6 & 1.56 & 0.70 & 1.06 & 1.29 & 76\\
14E3 & 199.6 & 1.53 & 0.69 & 1.06 & 1.29 & 77\\
14E4 & 201.0 & 1.63 & 0.77 & 1.18 & 1.45 & 69\\ \hline
\end{tabular}
\caption{Heliostat Sun spot sizes as measured from background subtracted
CCD images taken at Sandia.}
\end{center}
\end{table}

\newpage

\begin{table}
\begin{center}
\begin{tabular}{|lccc|} \hline
Timing Method  &   Small Pulseheights & Large Pulseheights & All Data\\
 & $\sigma_{t}$ & $\sigma_{t}$ & $\sigma_{t}$ \\
\hline \hline
Conventional TDCs     &    0.82 ns  &  0.87 ns  &   0.86 ns \\
Waveform Digitizers &    0.73 ns  &  0.51 ns  &   0.60 ns \\
\hline\end{tabular}
\caption{Comparison of measured timing resolution for conventional
TDCs and 1 GSample/sec waveform digitizers.  The waveform digitizers have
superior timing resolution at all pulseheights.}
\end{center}
\end{table}

\clearpage

\begin{figure}[!htp]
\vspace{8mm}
\subfigure[{\bf Figure 2.1:} The STACEE concept.
Cherenkov light produced in the air shower created by
an astrophysical $\gamma$-ray is beamed to the ground.
Solar heliostat mirrors reflect this light to a secondary
collector on the central tower which in turn reflects it
to a camera of photomultiplier tubes.]
{\psfig{figure=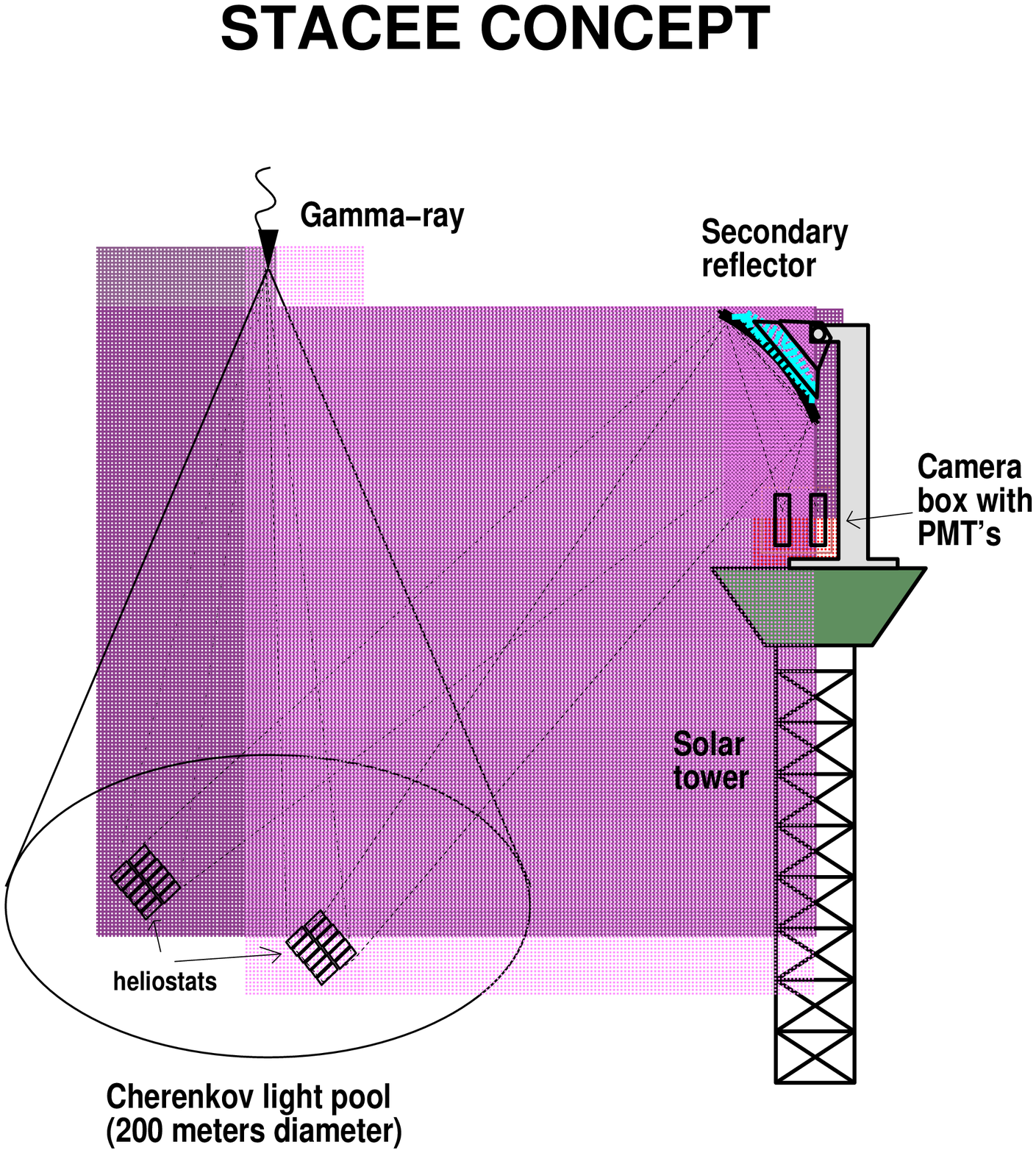,width=6in}}
\end{figure}

\begin{figure}[!htp]
\subfigure[{\bf Figure 2.2:}
STACEE telescope design, including secondary optic and camera.  The
STACEE instrument will consist of three such
telescope modules.  One of these has already been constructed and
tested successfully in the field.  Dimensions are given by the
scale on the right.]
{\psfig{figure=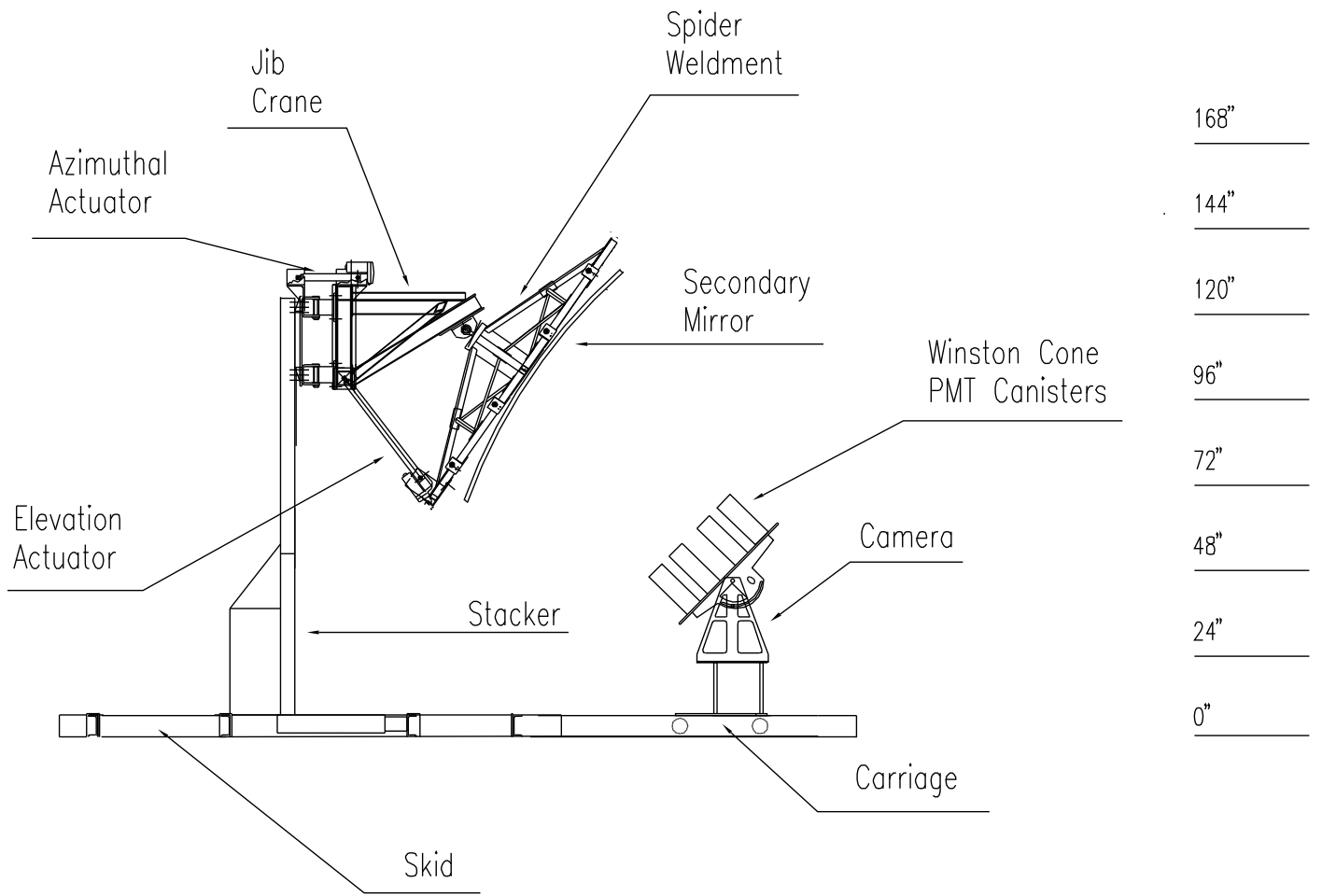,width=6in}}
\end{figure}

\clearpage 

\begin{figure}[!htp]
\subfigure[{\bf Figure 2.3:}
Schematic of the electronics setup used for the prototype tests
(one channel of eight).
Each PMT signal was AC-coupled and amplified.  Cable delays compensated
for the varying times-of-flight between heliostats and the central
tower.  Each signal was then split by a fan-out.  Analog copies of the
signal went to waveform digitizers and ADC units.  The signal was also
discriminated, and the discriminated outputs drove scaler units and
stoped TDCs.  The trigger was formed by combining
signals from all eight PMTs, and the trigger gated the waveform
digitizers and ADCs, and started the TDC units.]
{\psfig{figure=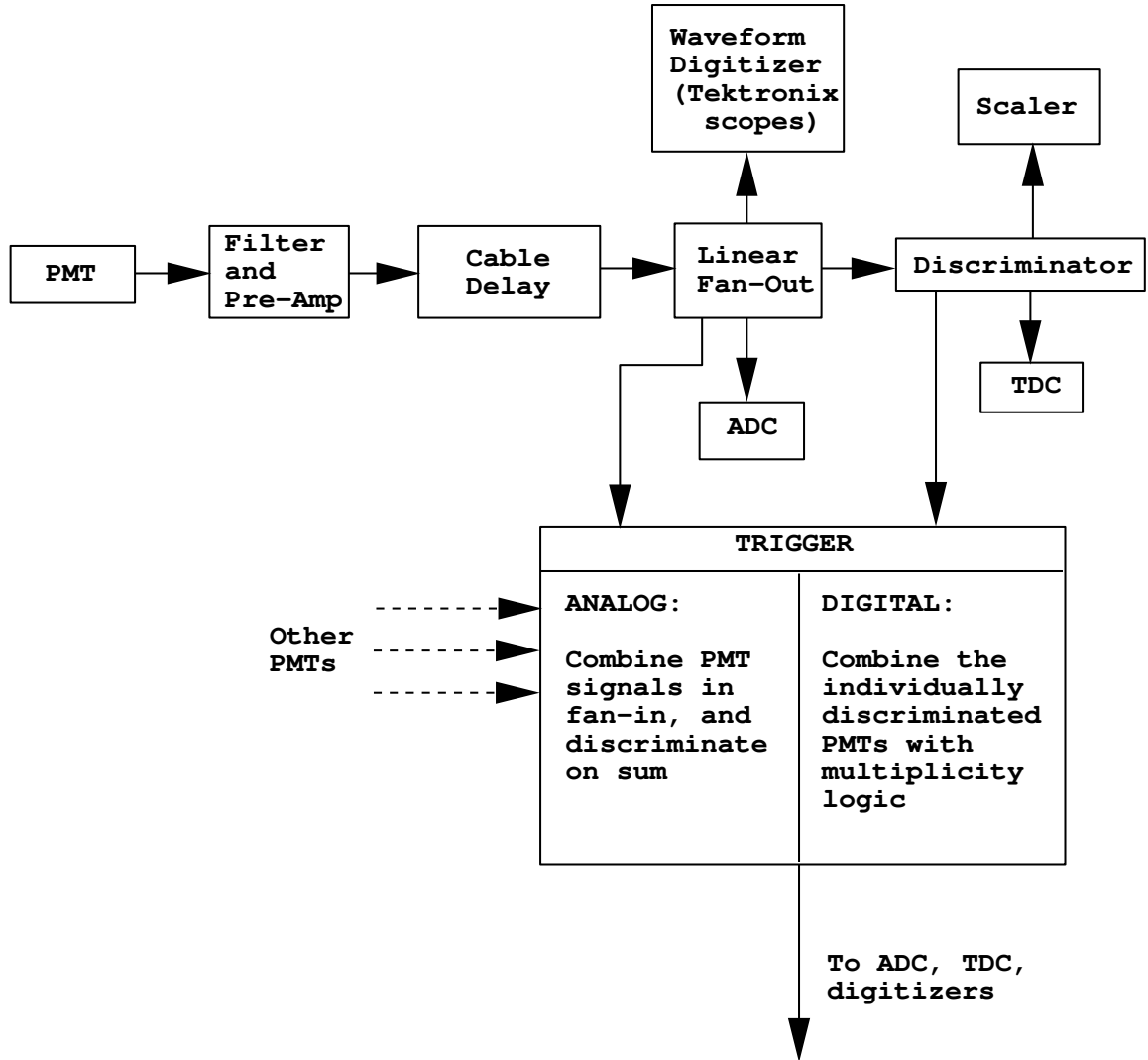,width=6in}}
\end{figure}

\begin{figure}[!htp]
\subfigure[{\bf Figure 2.4:}
Layout of the Sandia heliostat field.  Each square represents the location of a
single heliostat.  The sizes of the squares are not proportional to the physical
size of the heliostats.  The two different eight-heliostat 
configurations
used in the August and October tests are shown.]
{\psfig{figure=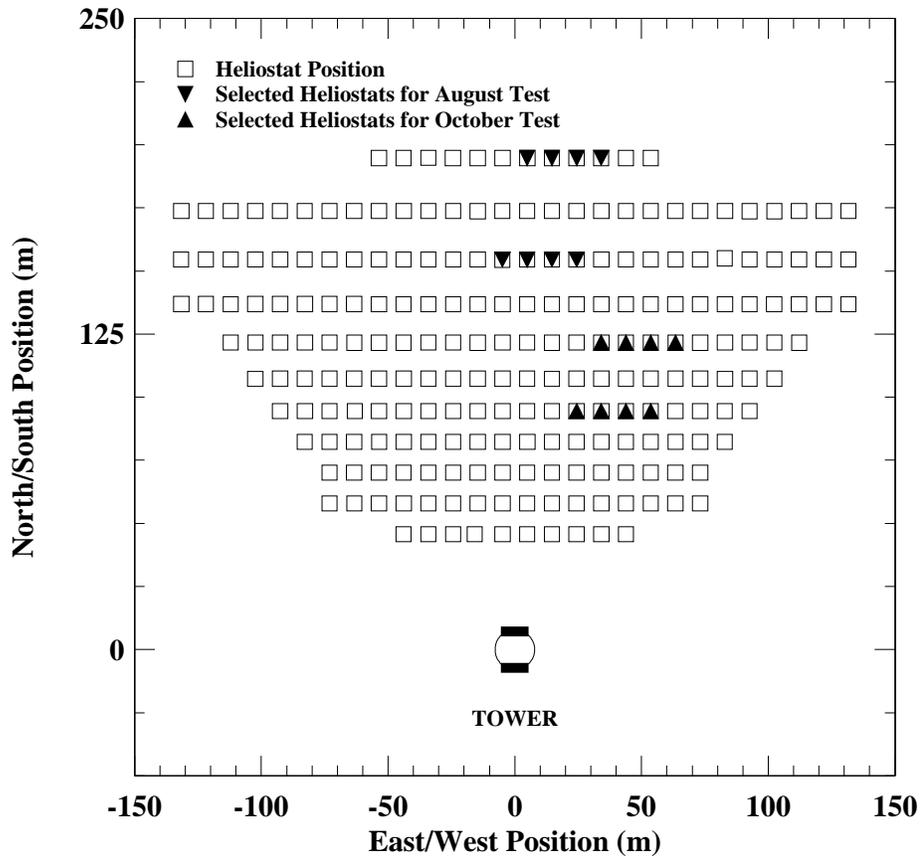,width=6in}}
\end{figure}

\begin{figure}[!htp]
\subfigure[{\bf Figure 3.1:}
Average number of hours of clear, moonless observing time expected per month.  
The expected fraction of clear skies is taken from daily meteorological records
for rain fall and percent sunshine for Albuquerque from 1948 to the present.
]
{\psfig{figure=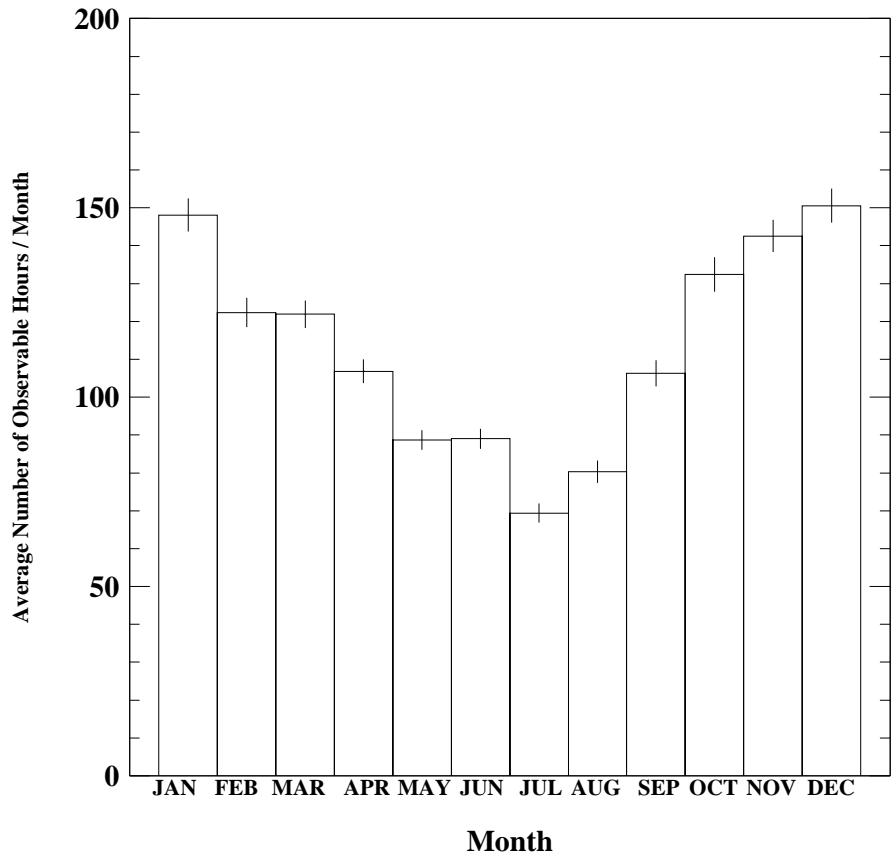,width=6in}}
\end{figure}

\begin{figure}[!htp]
\subfigure[{\bf Figure 3.2:}
Schematic drawing of the photometer setup used at Sandia.]
{\psfig{figure=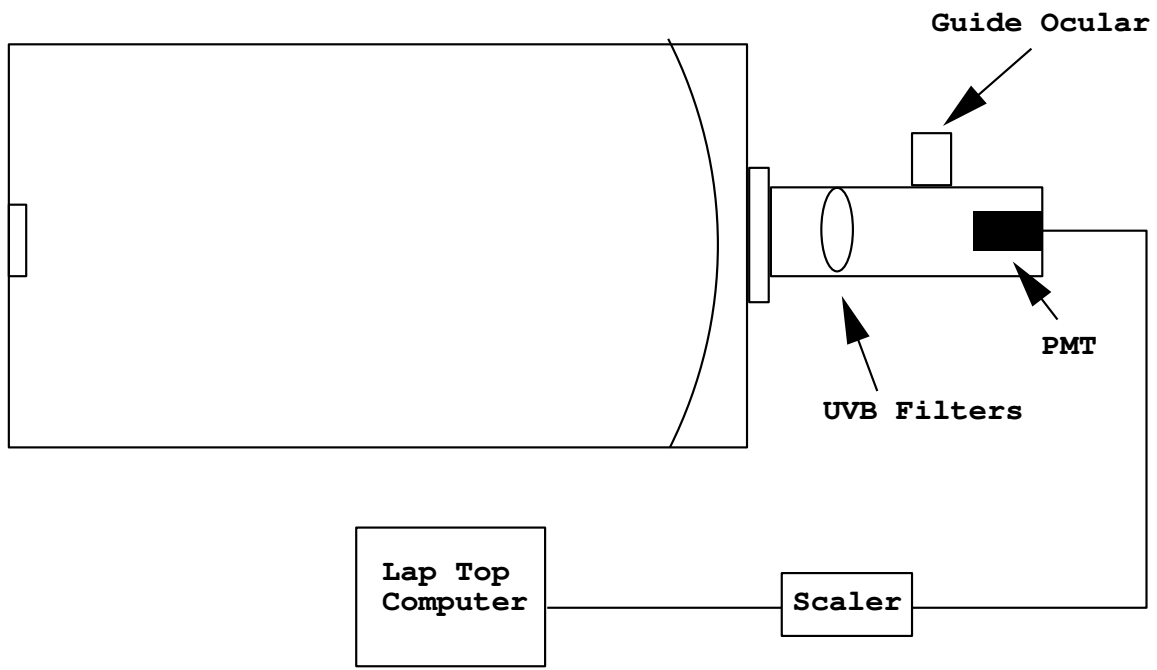,width=6in}}
\end{figure}

\begin{figure}[!htp]
\subfigure[{\bf Figure 3.3:}
Response curves for the standard photometric UBV filters \cite{allen}.]
{\psfig{figure=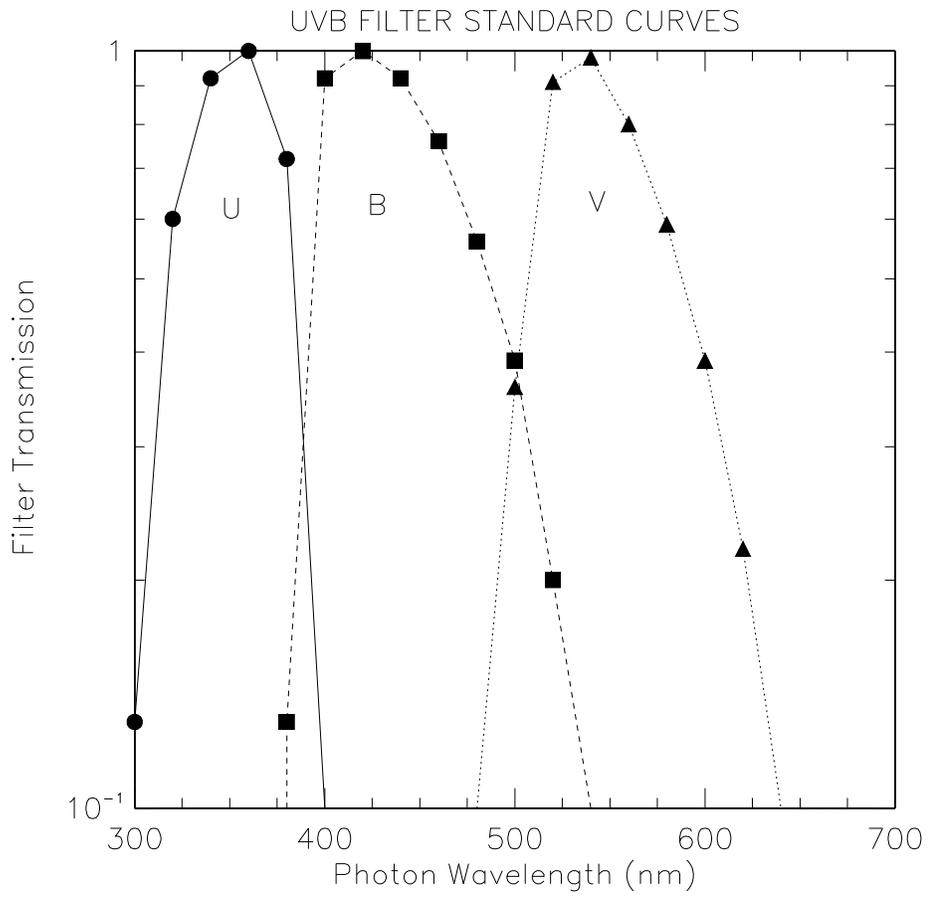,width=6in}}
\end{figure}

\begin{figure}[!htp]
\subfigure[{\bf Figure 3.4:}
Light curve for the star Mu Andromeda showing the PMT counting rate for
a single drift scan using the V-band filter.  Time bins are 0.1sec.]
{\psfig{figure=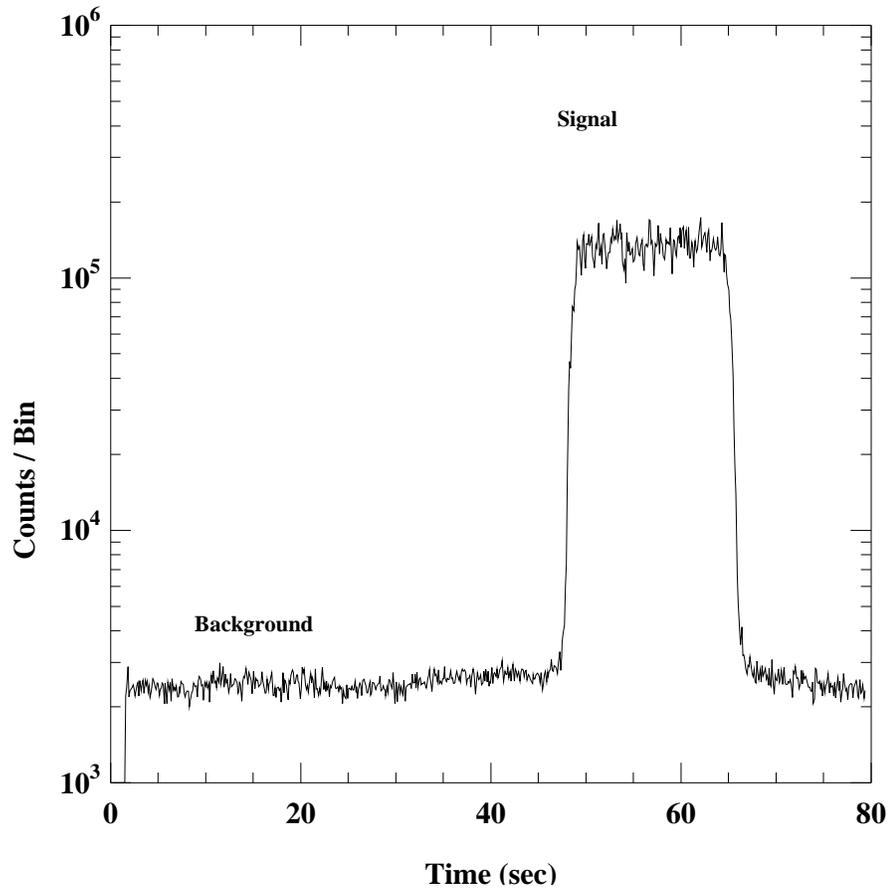,width=6in}}
\end{figure}

\begin{figure}[!htp]
\subfigure[{\bf Figure 3.5:}
Plot of stellar flux versus atmospheric depth for the star Mu Andromeda.]
{\psfig{figure=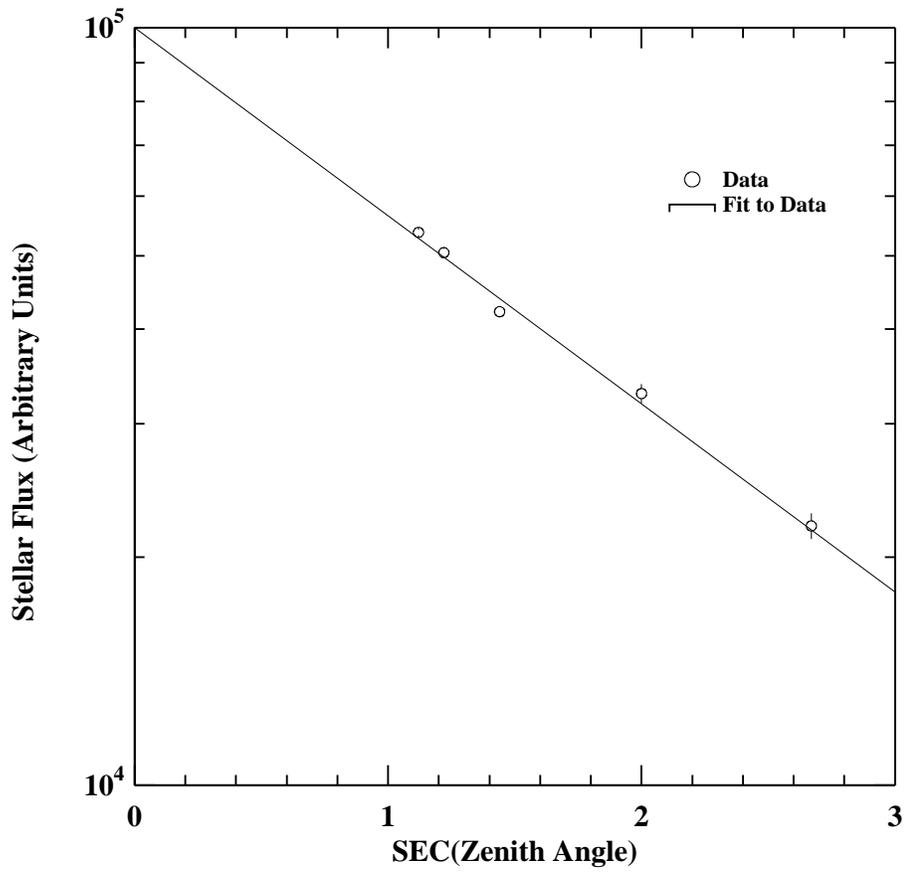,width=6in}}
\end{figure}

\begin{figure}[!htp]
\subfigure[{\bf Figure 3.6:}
Atmospheric transmission for the blue waveband, which contains most of
the observable Cherenkov light, as a function of zenith angle.  The points
and curves are identified in the legend.]
{\psfig{figure=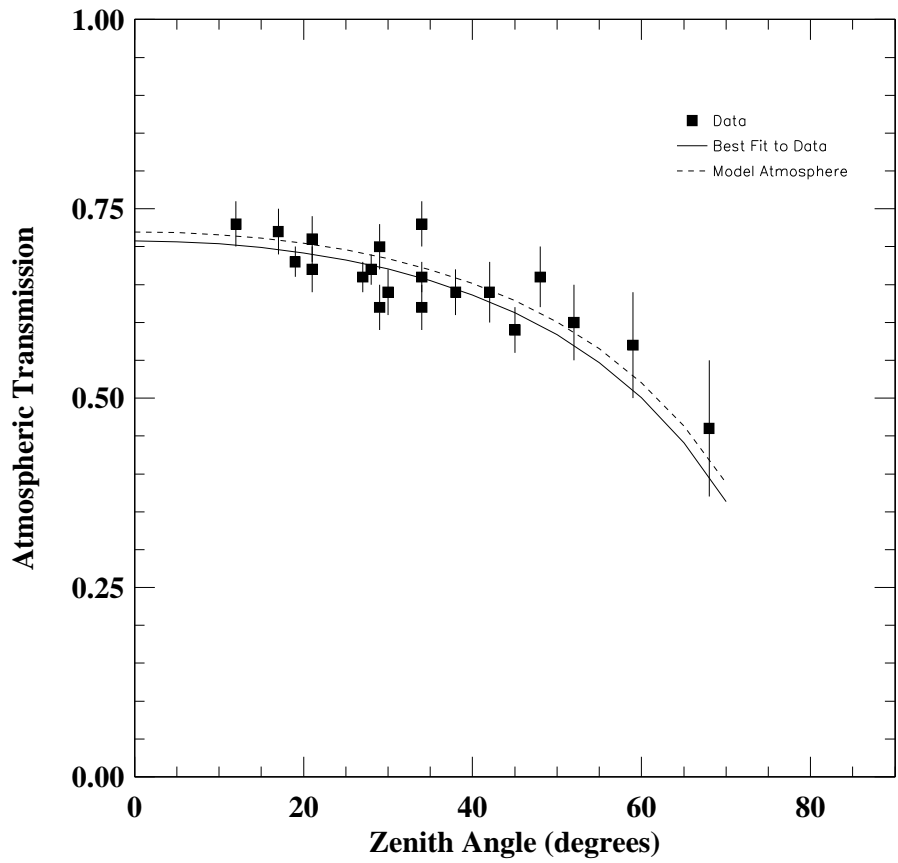,width=6in}}
\end{figure}

\begin{figure}[!htp]
\subfigure[{\bf Figure 3.7:}
Single photoelectron rates for the camera PMTs for two heliostat conditions:
edge facing the secondary, and viewing the night sky.  Note that the statistical 
errors on the data are smaller than the point sizes.]
{\psfig{figure=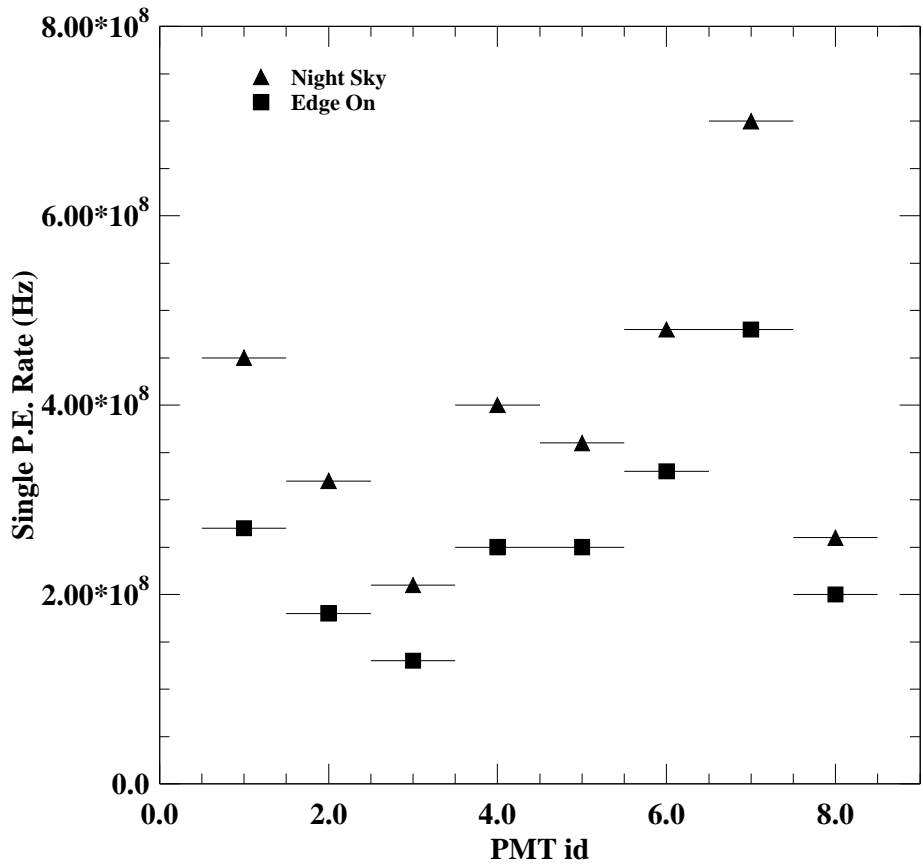,width=6in}}
\end{figure}

\begin{figure}[!htp]
\subfigure[{\bf Figure 4.1:}
Currents in two PMTs during a drift scan of the star Aldebaran.]
{\psfig{figure=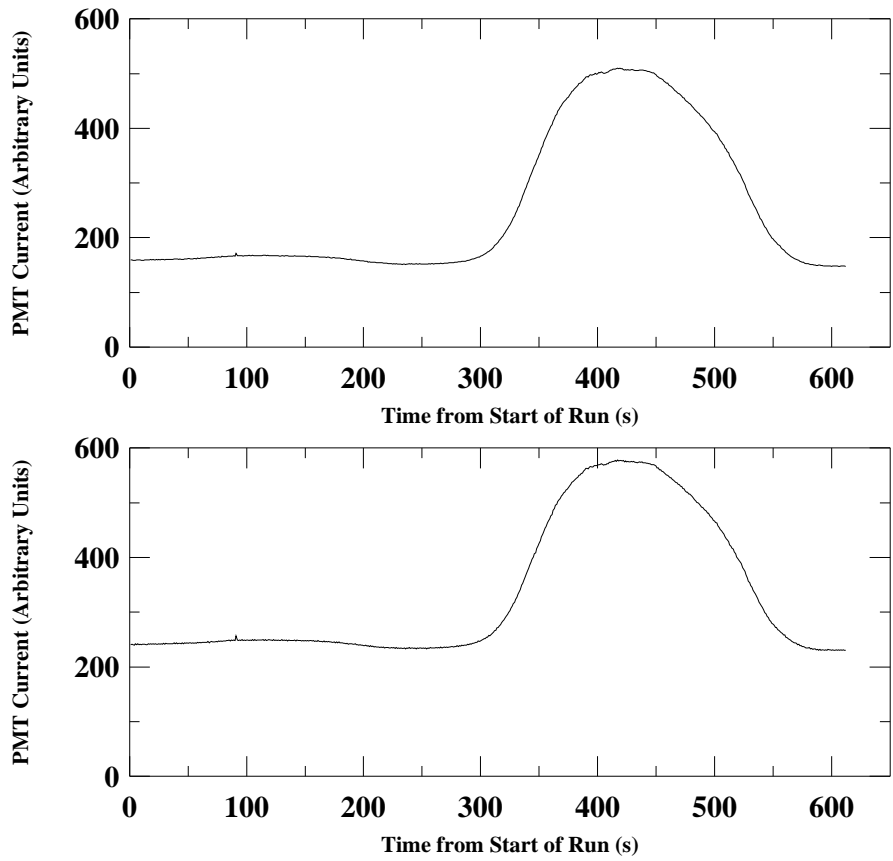,width=6in}}
\end{figure}

\begin{figure}[!htp]
\subfigure[{\bf Figure 4.2:}
PMT currents for two channels recorded while the
heliostats were tracking a bright star.  Note the suppressed zeroes on
the vertical scales.] 
{\psfig{figure=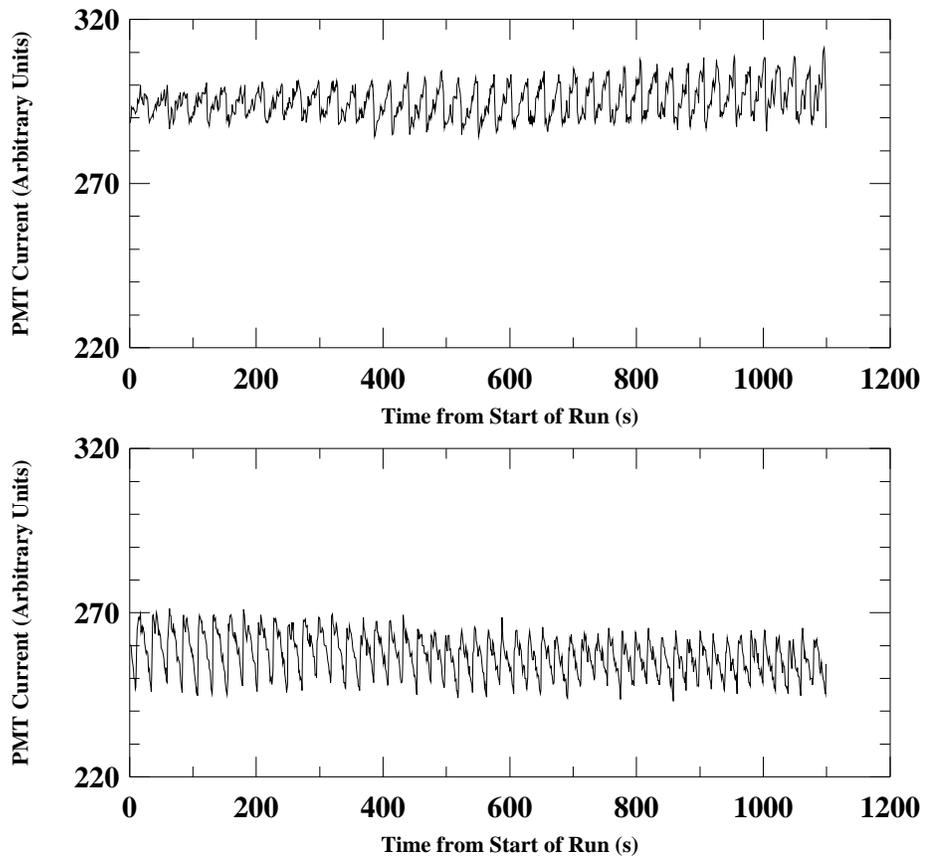,width=6in}}
\end{figure}

\begin{figure}[!htp]
\subfigure[{\bf Figure 4.3:}
CCD image of a single heliostat Sun spot projected onto the tower.  The
contours represent the fraction of the total light contained within a
given radius, starting at $10\%$
and increasing in steps of $20\%$.  The box underneath is a $1m
\times 1m$ box to indicate scale.]
{\psfig{figure=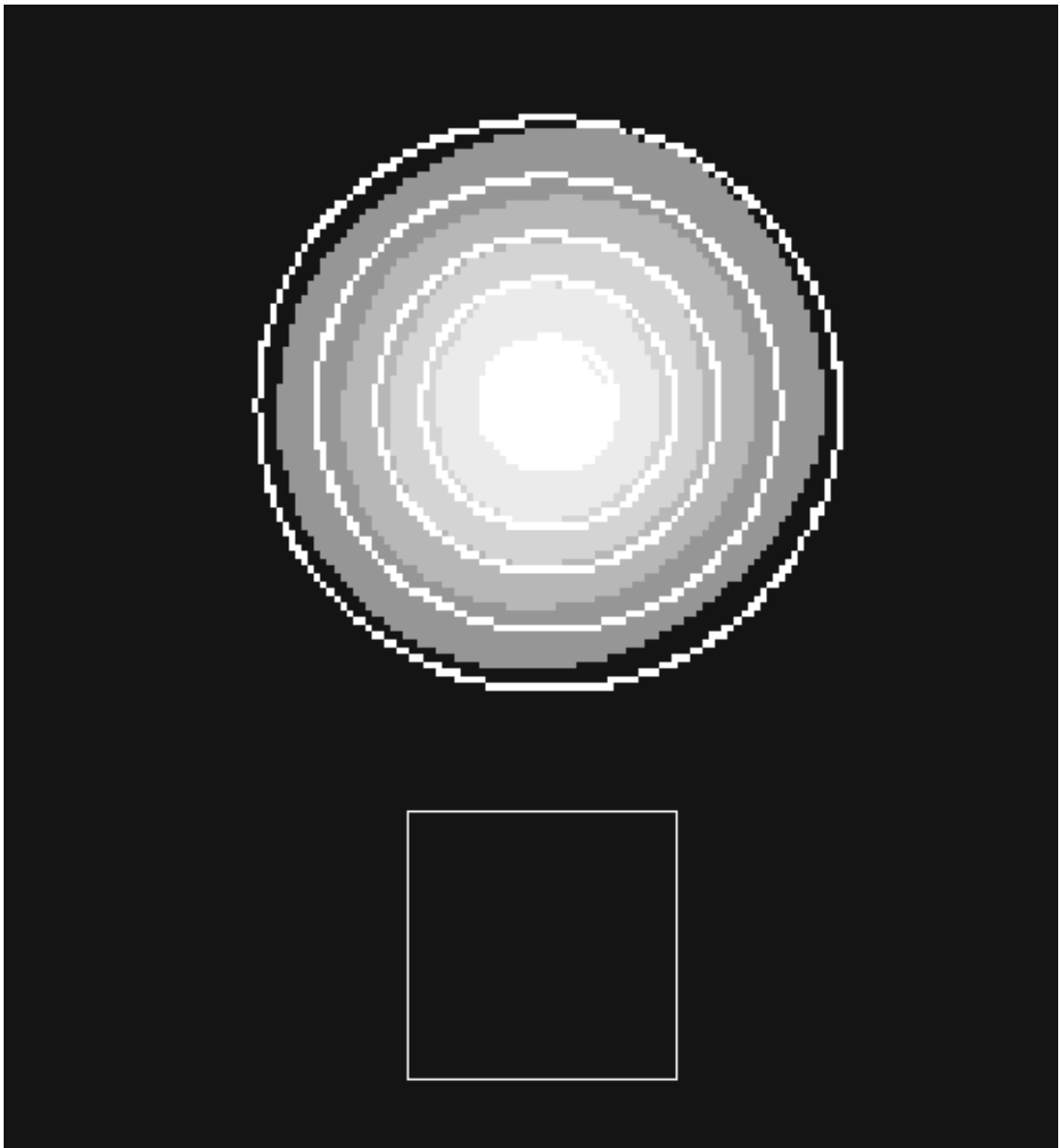,width=6in}}
\end{figure}

\clearpage

\begin{figure}[!htp]
\subfigure[{\bf Figure 4.4:}
Solar spot size (FWHM) projected on a target at the tower versus distance between
heliostat and target for 15 heliostats measured using a CCD camera.  Except for
one heliostat with particularly poor optical alignment (8E3 at $107\,$m)
the spots follow a regular trend, becoming less concentrated with greater
distance from the tower.]
{\psfig{figure=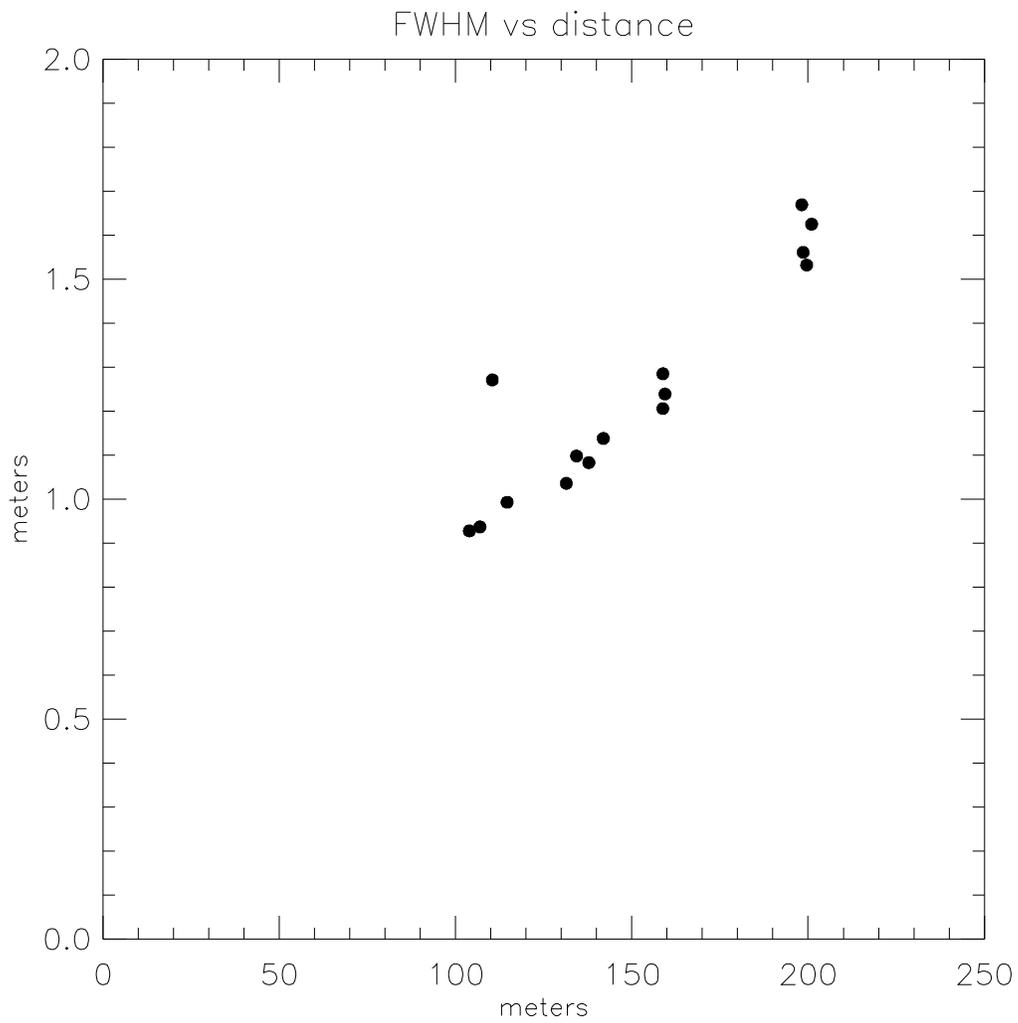,width=6in}}
\end{figure}

\begin{figure}[!htp]
\subfigure[{\bf Figure 4.5:}
Measured heliostat facet reflectivity, as a function of 
wavelength.]
{\psfig{figure=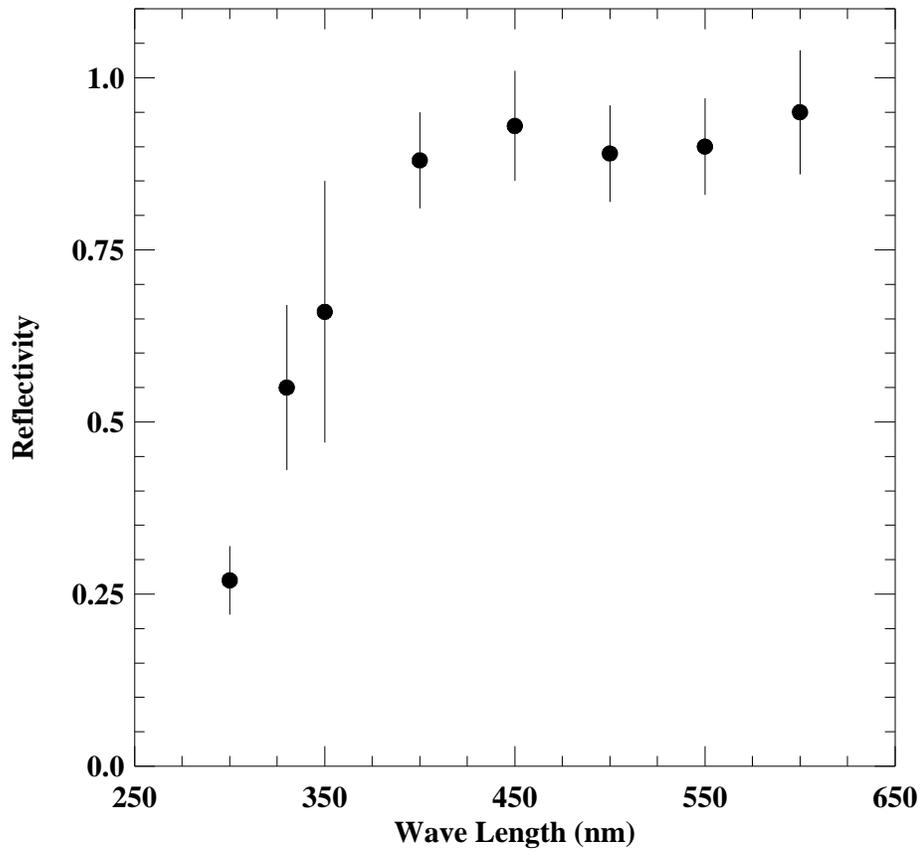,width=6in}}
\end{figure}

\begin{figure}[!htp]
\subfigure[{\bf Figure 5.1:}
A CCD image of the Sandia heliostat field projected onto a white lucite
placard at the focal plane of the
secondary used for the October 1996 observations.  This image was
obtained by viewing the placard at an oblique angle from the balcony
where the telescope mount is installed.  Grid lines drawn on the
placard represent a 1 cm spacing at the focal plane. Images of eight
selected heliostats are distinct and well-separated, with virtually no
optical overlap.]
{\psfig{figure=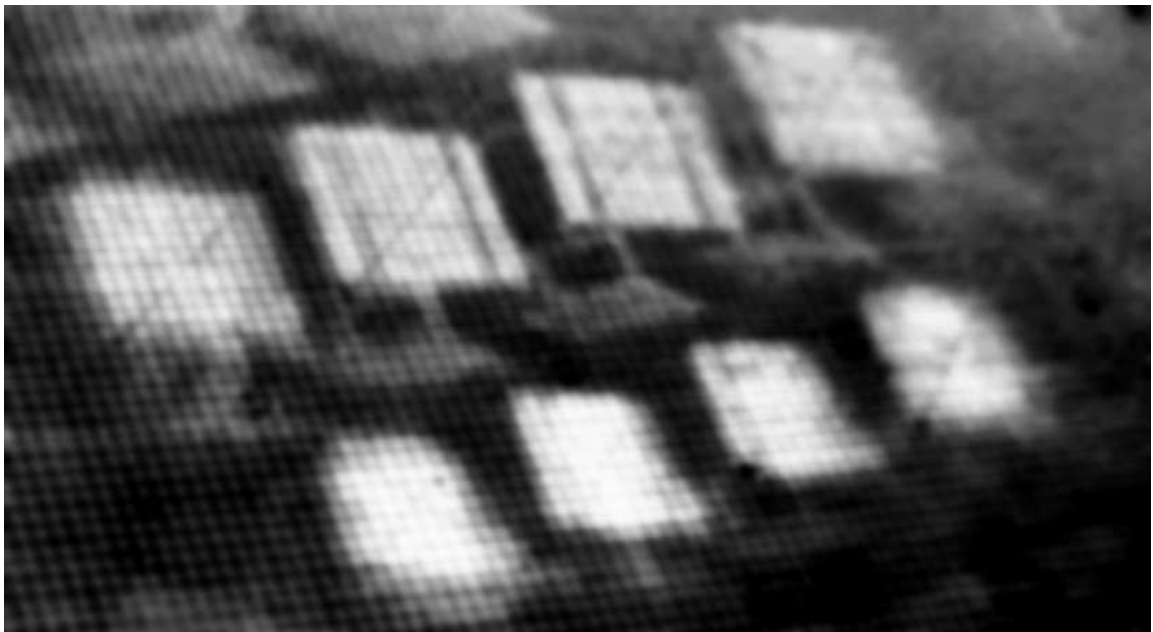,width=6in}}
\end{figure}

\begin{figure}[!htp]
\subfigure[{\bf Figure 5.2:}
Analog trigger rate as a function of discriminator threshold.
The data are compared to a model of single heliostat triggers from 
cosmic rays and muons, as identified
in the legend.
The heliostats were viewing separate parts of the sky.]
{\psfig{figure=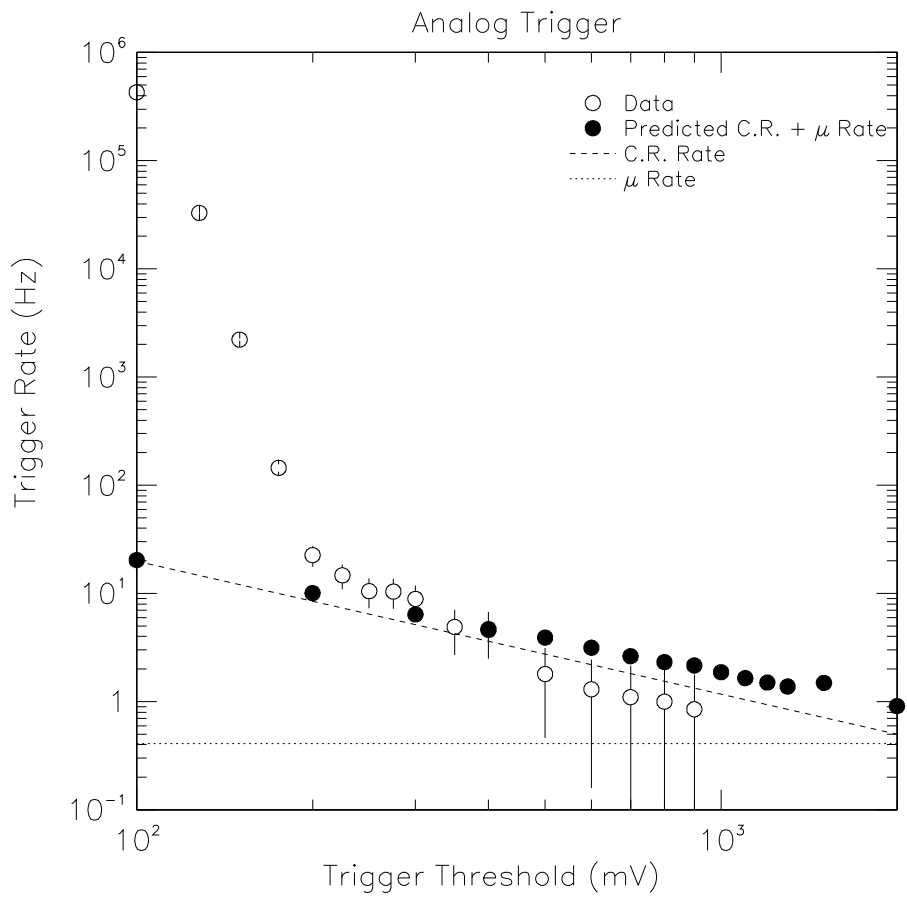,width=6in}}
\end{figure}

\begin{figure}[!htp]
\subfigure[{\bf Figure 5.3:}
Digital trigger rate as a function of individual channel discriminator
threshold.  The heliostats were viewing separate parts of the sky.  The
lines represent power law fits to the data, for different trigger 
configurations, as identified in the legend.]
{\psfig{figure=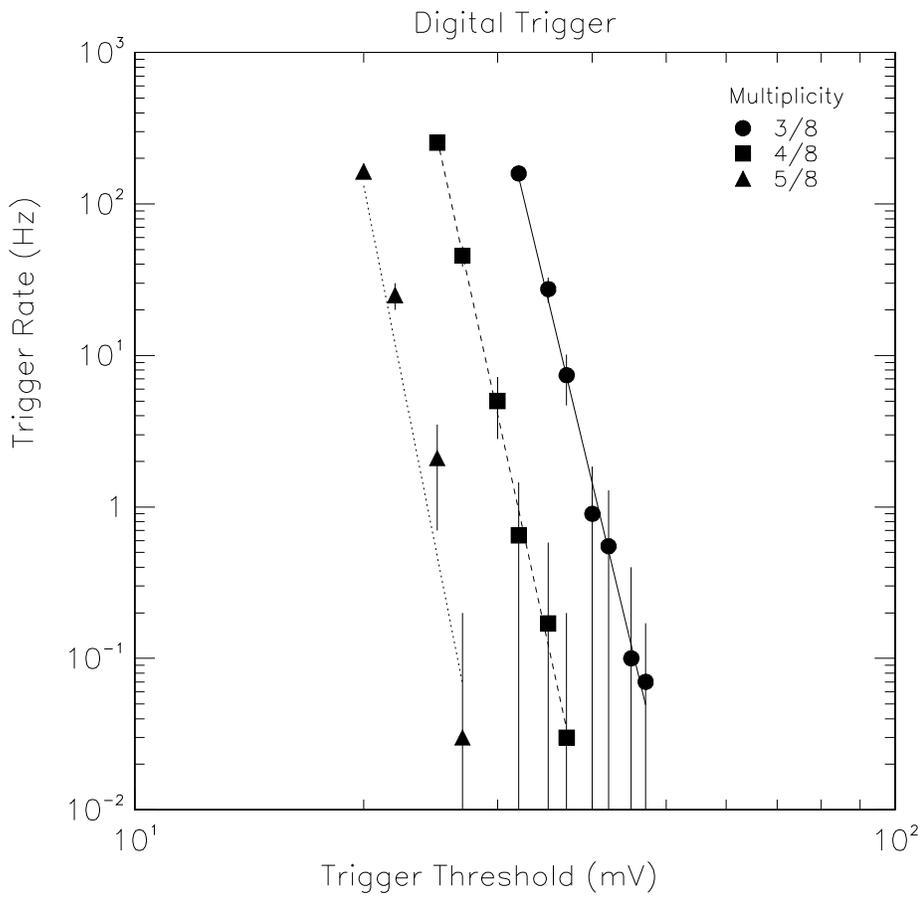,width=6in}}
\end{figure}

\begin{figure}[!htp]
\subfigure[{\bf Figure 5.4:}
Pulse height spectrum obtained from observations of air showers at the zenith.]
{\psfig{figure=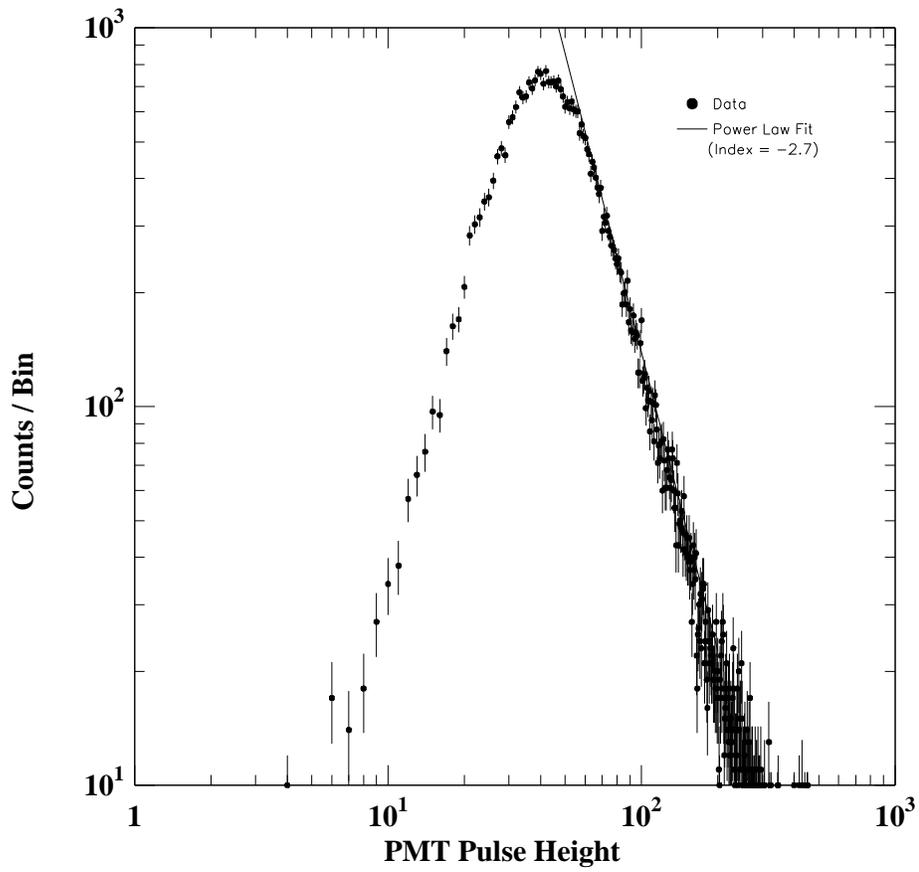,width=6in}}
\end{figure}

\begin{figure}[!htp]
\subfigure[{\bf Figure 5.5:}
Trigger rate as a function of zenith angle.]
{\psfig{figure=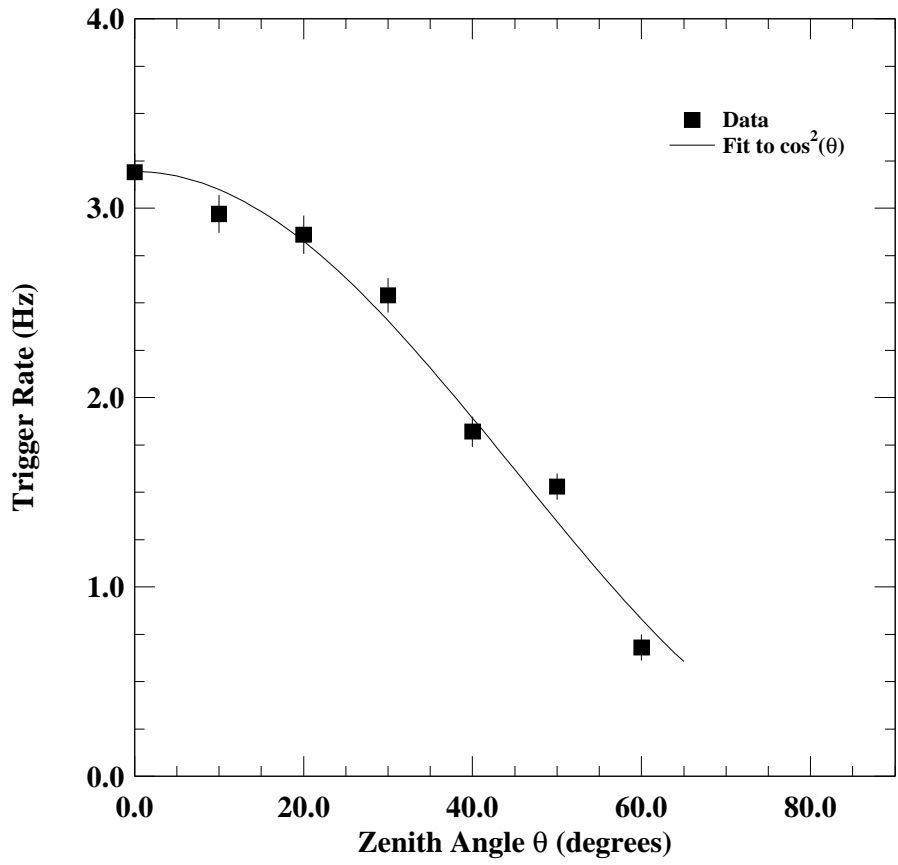,width=6in}}
\end{figure}

\begin{figure}[!htp]
\subfigure[{\bf Figure 5.6a:}
Diagram illustrating overlapping heliostat fields-of-view when the heliostats
are all aimed at a source at infinity.  Note that the entire track of the air
shower is not completely contained within the overlap of the heliostat 
fields-of-view.  Angles and distances have been exaggerated for purposes of
illustrating the concept.]
{\psfig{figure=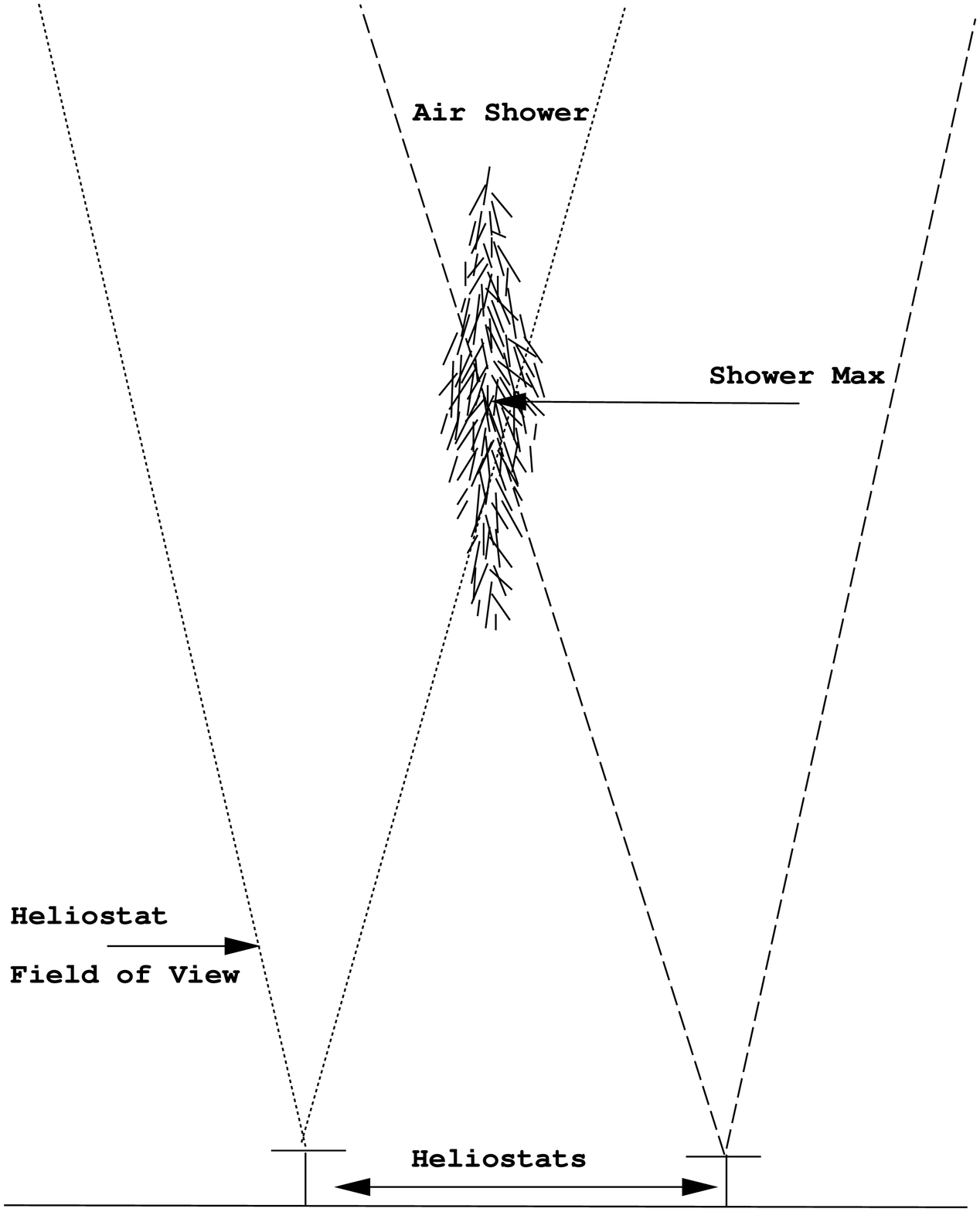,width=3in}}
\subfigure[{\bf Figure 5.6b:}
Diagram illustrating the advantage of convergent viewing.  When the heliostats
are canted in to view the interaction region the air shower is better contained
by the overlapping heliostat fields-of-view.]
{\psfig{figure=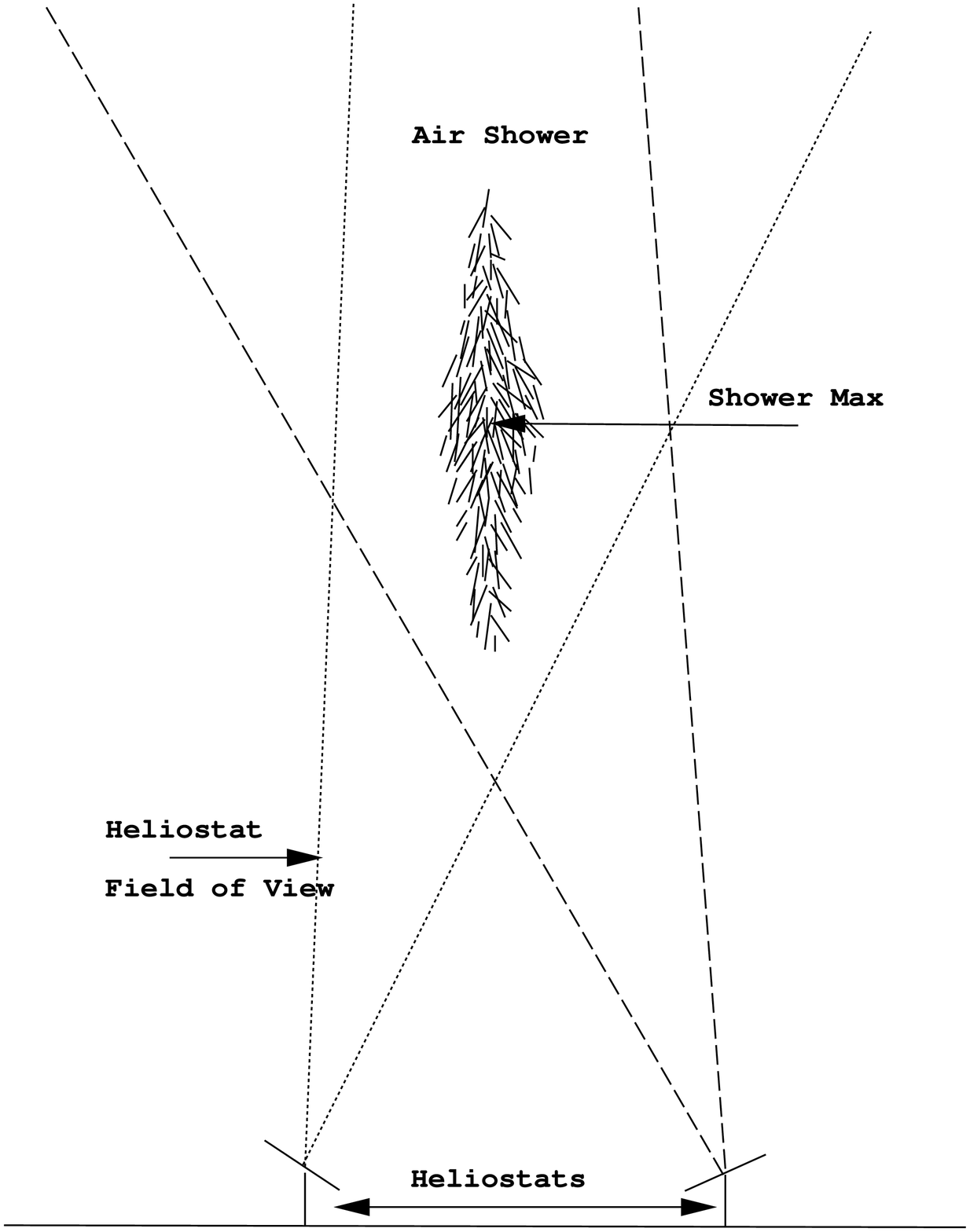,width=3in}}
\end{figure}

\begin{figure}[!htp]
\subfigure[{\bf Figure 5.7:}
Trigger rates as a function of heliostat canting angle.  A canting angle of
$0\degs$ corresponds to parallel viewing.  A 4/8 digital trigger was used
and the heliostats were viewing the zenith.]
{\psfig{figure=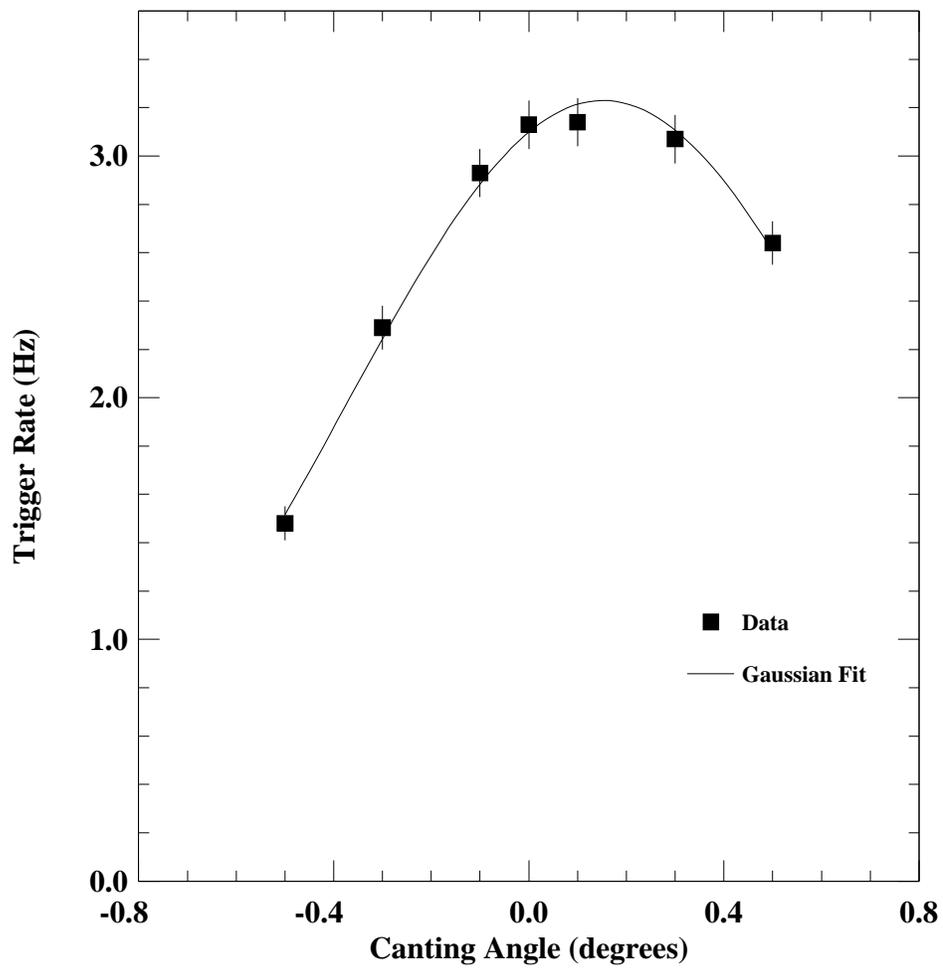,width=6in}}
\end{figure}

\begin{figure}[!htp]
\subfigure[{\bf Figure 5.8:}
The density of Cherenkov photons on the ground, within 100m of the
core, for vertically incident $\gamma$-ray and proton initiated air showers as
determined by the Hillas Monte Carlo.  The 
measured field of view of the Sandia heliostats has been accounted for
so that only those photons which would be collected by the STACEE secondary
telescope have been considered.
Note that the statistical errors are smaller that the size of the
symbols used in plotting the data.]
{\psfig{figure=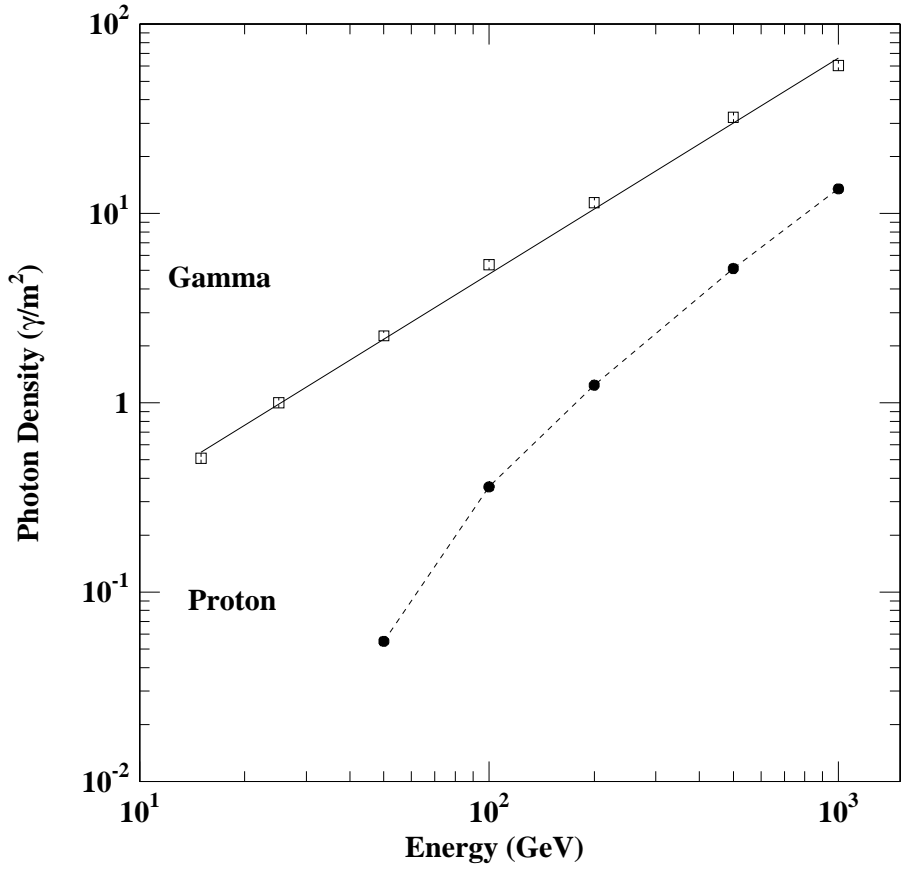,width=6in}}
\end{figure}

\end{document}